\documentclass[journal,onecolumn]{IEEEtran}

\usepackage{graphicx}
\usepackage{subfigure}
\usepackage{caption2}
\usepackage{amssymb}
\usepackage{amsthm}
\usepackage{amsmath}
\usepackage{array}
\usepackage{cite}

\begin{document}
%
\title{Degrees-of-Freedom Region of MISO-OFDMA Broadcast Channel with Imperfect CSIT}
%
%
%

\author{Chenxi Hao, Borzoo Rassouli and Bruno Clerckx 
\thanks{Chenxi Hao, Borzoo Rassouli and Bruno Clerckx are with the Communication and Signal Processing group of Department of Electrical and Electronics, Imperial College London, email: {chenxi.hao10;b.rassouli12;b.clerckx@imperial.ac.uk}}
\thanks{This paper was in part published in ''MISO Broadcast Channel with Imperfect and (Un)matched CSIT in the Frequency Domain: DoF Region and Transmission Strategies'', PIMRC'13.}
\thanks{This work was supported in part by Samsung Electronics and by the Seventh Framework Programme for Research of the European Commission under grant number HARP-318489.}}

\markboth{IEEE transactions on Information Theory}%
{Shell \MakeLowercase{\textit{et al.}}: Degrees-of-Freedom Region of MISO-OFDMA Broadcast Channel with Imperfect CSIT}

\maketitle

\begin{abstract}
This contribution investigates the Degrees-of-Freedom region of a two-user frequency correlated Multiple-Input-Single-Output (MISO) Broadcast Channel (BC) with imperfect Channel State Information at the transmitter (CSIT). We assume that the system consists of an arbitrary number of subbands, denoted as $L$. Besides, the CSIT state varies across users and subbands. A tight outer-bound is found as a function of the minimum average CSIT quality between the two users. Based on the CSIT states across the subbands, the \emph{DoF} region is interpreted as a weighted sum of the optimal \emph{DoF} regions in the scenarios where the CSIT of both users are perfect, alternatively perfect and not known. Inspired by the weighted-sum interpretation and identifying the benefit of the optimal scheme for the unmatched CSIT proposed by Chen et al., we also design a scheme achieving the upper-bound for the general $L$-subband scenario in frequency domain BC, thus showing the optimality of the \emph{DoF} region.
\end{abstract}

\IEEEpeerreviewmaketitle

\section{Introduction}\label{Intro}
Channel State Information at the Transmitter (CSIT) is crucial to the \emph{DoF} performance in downlink Broadcast Channel, but having perfect CSIT is a challenging issue. In practice, each user estimates, quantizes and reports its CSI to the transmitter. This process is subject to imperfectness and latency. Their impact on the \emph{DoF} region has attracted a lot of attention in recent years. The usefulness of the perfect but completely outdated CSIT was studied in \cite{Tse10}. Literature \cite{Tandon12} generalized the findings in \cite{Tse10} by giving an optimal \emph{DoF} region for an alternative CSIT setting, where the CSIT of each user can be perfect, delayed or none. Moreover, authors in \cite{Ges12,xinping_miso_conf} and \cite{Gou12} looked into the scenario with both perfect delayed CSIT and imperfect instantaneous CSIT, whose qualities are shown to make an impact on the optimal \emph{DoF} region. \cite{Chen12a} and \cite{CB12} extended the results of \cite{Ges12} and \cite{Gou12} by considering the different qualities of instantaneous CSIT of the two users. Furthermore, the authors of \cite{Chen12b} studied the scenario where both delayed CSIT and instantaneous CSIT are imperfect and the results were generalized to a scenario with multiple slots and evolving CSIT states in \cite{Jinyuan_evolving_misobc}. Recently, all the results found in two-user time-correlated MISO BC with delayed CSIT have been extended to the MIMO case in \cite{xinping_mimo,xinping_mimo_conf,Jinyuan_BCIC}. Other works, such as \cite{VV11a,VV11b,HJSV09,Lee,TJS12,xinping_finiteSNR,xinping_Kuser,Jinyuan_diversity,Kerret_APZF} have covered other related topics about the \emph{DoF} region of time domain BC.

However, in practical systems like Long Term Evolution (LTE), the system performance loss of Multiuser MIMO (MU-MIMO) is primarily due to CSI measurement and feedback inaccuracy rather than delay \cite{BrunoBook}. Therefore, assuming the CSI arrives at the transmitter instantaneously, we are interested in the frequency domain BC where the CSI is measured and reported to the transmitter on a per-subband basis. Due to frequency selectivity, constraints on uplink overhead and user distribution in the cell, the quality of CSI reported to the transmitter varies across users and subbands.

The alternating CSIT state ($I_{11}I_{12}{=}NP$ and $I_{21}I_{22}{=}PN$\footnote{$I_{jk}$ is the CSIT state of user $k$ in subband $j$. $PP$ means perfect CSIT for both users; $NN$ stands for no CSIT for both users; $PN/NP$ refer to the CSIT states alternating between Perfect/None and None/Perfect.}) can be interpreted as two users reporting their CSI in two different subbands. Those unmatched CSIT was shown still useful in benefiting the \emph{DoF} region in \cite{Tandon12}. A sum \emph{DoF} of $\frac{3}{2}$ is achieved, outperforming that in the case without CSIT. The scheme proposed in \cite{Tandon12}, called $S_3^{3/2}$, transmits two private symbols and one common message (to be decoded by both users) in two channel uses (subbands/slots). The key point lies in sending the common message twice in different subbands, so that the two users can decode it in turn due to the alternating CSIT in each subband. With the knowledge of the common message, the private symbols are recovered.

A more general scenario consists in having the channel state changing to $I_{11}I_{12}{=}\beta\alpha$ and $I_{21}I_{22}{=}\alpha\beta$ (where $\alpha$ and $\beta$ represent the quality of the imperfect CSIT, both ranging from $0$ to $1$). Literature \cite{icc13freq} was the first work investigating this issue. A novel transmission strategy integrating Maddah-Ali and Tse (MAT) scheme, ZFBF and FDMA was proposed. Recently, the \emph{DoF} region found in \cite{icc13freq} has been improved by the scheme proposed in \cite{Elia13}, which combines $S_3^{3/2}$ scheme, ZFBF and FDMA. It outperforms \cite{icc13freq} because no extra channel use is required to decode all the symbols. The \emph{DoF} region in the alternating $(\alpha{,}\beta)$ scenario has been conversed in our conference paper \cite{pimrc2013}. The optimal \emph{DoF} region was interpreted as a weighted sum of the \emph{DoF} region in the CSIT state $PP$, $PN/NP$ and $NN$. The weights are functions of the CSIT qualities of the two users, revealing an equivalence between the CSIT quality and the fraction of time when the CSIT is perfect as in \cite{Tandon12}.

So far, the literature addressing the problem of frequency domain BC (or time domain BC without delayed CSIT) focuses on two subbands and assumes that the CSIT states alternate. This assumption is relatively optimistic as in a more realistic wireless communication framework two users may be scheduled simultaneously on multiple subbands. The channels in different subbands may have weak correlation due to the frequency selectivity. The qualities of the CSIT can also vary across users and subbands. We aim at understanding whether the multiple and arbitrary CSIT state can synergistically boost the \emph{DoF} region. In this paper, we generalize our results of \cite{pimrc2013} to an $L$-subband scenario with arbitrary values of the CSIT qualities of both users (see Figure \ref{fig:system}). In particular, we highlight the main contributions as follows:
\begin{enumerate}
\item We derive a tight outer-bound to the \emph{DoF} region in the $L$-subband frequency correlated MISO BC with arbitrary values of CSIT qualities. It is shown to be a function of the minimum average CSIT quality between the two users. The converse relies on the upper-bound in \cite{NairGamal}, the extremal inequality \cite{Extremal} and Lemma 1 in \cite{Ges12}.
\item The \emph{DoF} region is interpreted as the weighted sum of the \emph{DoF} region in the subchannels with state $PP$, $PN/NP$ and $NN$, after we decompose the subbands into subchannels according to the qualities of the imperfect CSIT. The weights refer to the fraction of channel use of each type of the subchannels. For a given average CSIT quality but different distributions of the quality in each subband, we find the \emph{DoF} region remains unchanged but the compositions of the region are varying. Besides, we find a similar expression of the \emph{DoF} region as in \cite{Tandon12}, if we interpret the average CSIT quality as the fraction of channel use where the CSIT is perfect. This weighted-sum interpretation also provides an instructive insight into the achievable scheme.
\item By identifying the sub-optimality in the scheme proposed in \cite{icc13freq} and the optimality of the scheme in \cite{Elia13} for a 2-subband scenario, we propose the optimal transmission strategy achieving the outer-bound of the \emph{DoF} region in a $3$-subband scenario with $\sum_{j{=}1}^3a_j{=}\sum_{j{=}1}^3b_j$ ($a_j$ and $b_j$ are the qualities of user 1 and user 2 respectively in subband $j$). Also, we extend this scheme to the $L$-subband scenario with $\sum_{j{=}1}^La_j{=}\sum_{j{=}1}^Lb_j$. The key point lies in generating multiple common messages and sending them twice such that the two users can recover them alternatively and decode the private symbols afterwards.
\item Following the footsteps of the construction of the optimal scheme in the case with $\sum_{j{=}1}^La_j{=}\sum_{j{=}1}^Lb_j$, we design an optimal transmission strategy for the $L$-subband scenario with $\sum_{j{=}1}^La_j{\neq}\sum_{j{=}1}^Lb_j$.
\end{enumerate}

The rest of this paper is organized as follows. The system model is introduced in Section \ref{SMandMR}, where the main results are also included. The converse of the \emph{DoF} region is provided in Section \ref{conv}. A weighted-sum interpretation of the optimal \emph{DoF} region is derived in Section \ref{wsum}. In Section \ref{achp}, by analyzing the achievability in the two-subband scenario, the optimal transmission strategy for $L$-subband with $\sum_{j{=}1}^La_j{=}\sum_{j{=}1}^Lb_j$ is designed. In Section \ref{achq}, we build the transmission strategy for $L$-subband with $\sum_{j{=}1}^La_j{\neq}\sum_{j{=}1}^Lb_j$. Section \ref{conclusion} concludes the paper.

The following notations are used throughout the paper. Bold lower case letters stand for vectors whereas a symbol not in bold font represents a scalar. $\left({\cdot}\right)^T$ and $\left({\cdot}\right)^H$ represent the transpose and conjugate transpose of a matrix or vector respectively. $\mathbf{h}^\bot$ denotes the orthogonal space of the channel vector $\mathbf{h}$. $\mathcal{E}\left[{\cdot}\right]$ refers to the expectation of a random variable, vector or matrix. $\parallel{\cdot}\parallel$ is the norm of a vector. $A_{j_1}^{j_2}$ refers to the set $\{A_{j_1}{,}A_{j_1{+}1}{,}\cdots{,}A_{j_2}\}$, if $j_1{\leq}j_2$, otherwise $A_{j_1}^{j_2}{=}\emptyset$. $|A_{j_1}^{j_2}|$ represents the cardinality of set $A_{j_1}^{j_2}$, which equals to $j_2{-}j_1$. If $a$ is a scalar, $|a|$ is the absolute value of $a$. $f\left(P\right){\sim}{P^{B}}$ corresponds to $\lim \limits_{P{\to}{\infty}}\frac{{\log}f\left(P\right)}{{\log}P}{=}B$, where $P$ is SNR throughout the paper and logarithms are in base $2$. 

\section{System Model and Main Results}\label{SMandMR}
\subsection{Frequency domain two-user MISO BC}
\begin{figure}[t]
\renewcommand{\captionfont}{\small}
\centering
\includegraphics[height=5cm,width=16cm]{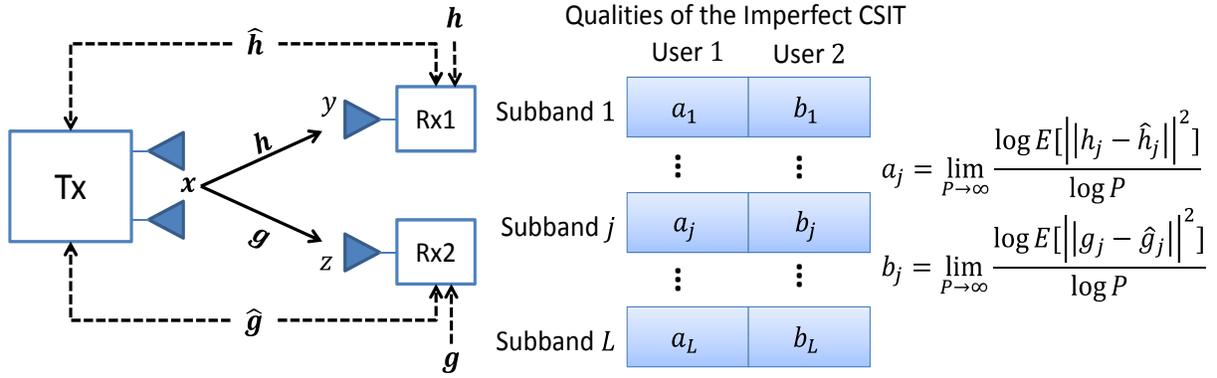}
\caption{System model of two-user MISO Broadcast Channel, with arbitrary values of the CSIT qualities across $L$ subbands.}\label{fig:system}
\end{figure}

In this contribution, we consider a system as shown in Figure \ref{fig:system}, which has a transmitter with two antennas and two users each with a single antenna. Denoting the transmit signal as $\mathbf{x}_j$ subject to $E[||\mathbf{x}_j||^2]{\leq}P$, the observations at user 1 and 2, $y_j$ and $z_j$ respectively, are given by
\begin{align}
y_j{=}&\mathbf{h}_j^H\mathbf{x}_j{+}\epsilon_{j1},\\
z_j{=}&\mathbf{g}_j^H\mathbf{x}_j{+}\epsilon_{j2},
\end{align}
where $j{\in}[1{,}L]$. $\epsilon_{j1}$ and $\epsilon_{j2}$ are unit power AWGN noise. $\mathbf{h}_j$ and $\mathbf{g}_j$, both with unit norm, are respectively the CSI of user 1 and user 2 in \emph{subband $j$}. The CSI are i.i.d across users and subbands. In this contribution, the transmit signal can be made up of three kinds of messages:
\begin{itemize}
  \item \emph{Common message I}, denoted as $c_j$ hereafter, is broadcast to both users in subband $j$. They should be recovered by both users, but can be intended exclusively for user 1 or user 2;
  \item \emph{Common message II}, denoted as $u_0(\cdot)$ hereafter, should be recovered by both users, but can be intended exclusively for user 1 or user 2. Unlike $c_j$, $u_0(\cdot)$ is broadcast twice, i.e. once in the subbands where the quality of CSIT of user 1 is higher than that of user 2, and once in the subbands where the quality of CSIT of user 2 is higher than that of user 1;
  \item \emph{Private message}, is intended for one user only, namely $u_j$ for user 1 and $v_j$ for user 2 in subband $j$.
\end{itemize}

\subsection{CSI Feedback Model}
Classically, in Frequency Division Duplexing (FDD), each user estimates their CSI in the specified subband using pilot and the estimated CSI is quantized and reported to the transmitter via a rate-limited link. In Time Division Duplexing (TDD), CSI is measured on the uplink and used in the downlink assuming channel reciprocity. We assume a general setup where the transmitter obtains the CSI instantaneously, but with imperfectness, due to the estimation error and/or finite rate in the feedback link.

Denoting $\hat{\mathbf{h}}_j$ and $\hat{\mathbf{g}}_j$ as the imperfect CSI of user 1 and user 2 in \emph{subband $j$} respectively, the CSI of user 1 and user 2 can be respectively modeled as
\begin{equation}
\mathbf{h}_j{=}\hat{\mathbf{h}}_j{+}\tilde{\mathbf{h}}_j,\quad \mathbf{g}_j{=}\hat{\mathbf{g}}_j{+}\tilde{\mathbf{g}}_j,\quad j{=}1{\cdots}L,
\end{equation}
where $\tilde{\mathbf{h}}_j$ and $\tilde{\mathbf{g}}_j$ are the corresponding error vectors, respectively with the covariance matrix $\mathbb{E}[\tilde{\mathbf{h}}_j\tilde{\mathbf{h}}_j^H]{=}\sigma_{j1}^2\mathbf{I}_2$ and $\mathbb{E}[\tilde{\mathbf{g}}_j\tilde{\mathbf{g}}_j^H]{=}\sigma_{j2}^2\mathbf{I}_2$. $\hat{\mathbf{h}}_j$ and $\hat{\mathbf{g}}_j$ are respectively independent of $\tilde{\mathbf{h}}_j$ and $\tilde{\mathbf{g}}_j$. The norm of $\hat{\mathbf{h}}_j$ and $\hat{\mathbf{g}}_j$ scale as $P^0$ at infinite SNR.

We employ the notation $\mathcal{S}_j{\triangleq}\{\mathbf{h}_j{,}\mathbf{g}_j\}$ to represent the CSI of both users in subband $j$. Similarly, $\hat{\mathcal{S}}_j{\triangleq}\{\hat{\mathbf{h}}_j{,}\hat{\mathbf{g}}_j\}$ is the set of the imperfect CSI, $\tilde{\mathcal{S}}_j{\triangleq}\{\tilde{\mathbf{h}}_j{,}\tilde{\mathbf{g}}_j\}$ refers to the set of the CSI errors and $\mathcal{S}_j{=}\{\hat{\mathcal{S}}_j{,}\tilde{\mathcal{S}}_j\}$. $\hat{\mathcal{H}}_1^n$ and $\tilde{\mathcal{H}}_1^n$ respectively refer to sets of the imperfect CSI and the CSI error of user 1 while $\hat{\mathcal{G}}_1^n$ and $\tilde{\mathcal{G}}_1^n$ are similarly defined. In addition, $\hat{\mathcal{S}}_1^n$ is available at both the transmitter side and the receiver side. $\tilde{\mathcal{H}}_1^n$ and $\tilde{\mathcal{G}}_1^n$ are only perfectly known by user 1 and user 2 respectively.

To investigate the impact of the imperfect CSIT on the \emph{DoF} region, we assume that the variance of each entry in the error vector exponentially scales with SNR as in \cite{Ges12,xinping_miso_conf,Chen12a,CB12,Chen12b,Jinyuan_evolving_misobc,xinping_mimo,
xinping_mimo_conf,Jinyuan_BCIC,xinping_Kuser,Kerret_APZF,Carie10,icc13freq,Elia13,pimrc2013}, namely $\sigma_{j1}^2{\sim}P^{-a_j}$ and $\sigma_{j2}^2{\sim}P^{-b_j}$. $a_j$ and $b_j$ are respectively interpreted as the quality of the CSIT of user 1 and user 2 in subband $j$, given as follows
\begin{equation}
a_j{=}\lim_{P{\to}\infty}{-}\frac{{\log}\sigma_{j1}^2}{{\log}P}{,}\quad b_j{=}\lim_{P{\to}\infty}{-}\frac{{\log}\sigma_{j2}^2}{{\log}P}.
\end{equation}
$a_j$ and $b_j$ vary within the range of $\left[0{,}1\right]$. $a_j{=}1$ (resp. $b_j{=}1$) is equivalent to perfect CSIT because the full \emph{DoF} region can be achieved by simply doing ZFBF. $a_j{=}0$ (resp. $b_j{=}0$) is equivalent to no CSIT because it means that the variance of the CSI error scales as $P^0$, such that the imperfect CSIT cannot benefit the \emph{DoF} when doing ZFBF. Besides, $a_j$ and $b_j$ vary across all the $L$ subbands. It is important to note the following quantities
\begin{align}
\mathcal{E}[|\mathbf{h}_j^H\hat{\mathbf{h}}_j^\bot|^2]{=}&\mathcal{E}[|(\hat{\mathbf{h}}_j{+}\tilde{\mathbf{h}}_j)^H\hat{\mathbf{h}}_j^\bot|^2]\\
{=}&\mathcal{E}[|\tilde{\mathbf{h}}_j^H\hat{\mathbf{h}}_j^\bot|^2]\\
{=}&\mathcal{E}[\tilde{\mathbf{h}}_j^H\hat{\mathbf{h}}_j^\bot\hat{\mathbf{h}}_j^{\bot H}\tilde{\mathbf{h}}_j]{\sim}P^{-a_j}.
\end{align}
as they are frequently used in the achievable schemes in Section \ref{achp} and \ref{achq}. Similarly, we have $\mathcal{E}[|\mathbf{g}_j^H\hat{\mathbf{g}}_j^\bot|^2]{\sim}P^{-b_j}$.

It is worth noting that the CSIT pattern in Figure \ref{fig:system} is applicable to time domain. Specifically, the CSI report from each user arrives at the transmitter without latency, but it is imperfect due to the estimation error and/or finite rate in the feedback link. As the location of the users and their channel condition changes with time, the CSIT quality varies across users and transmission time-slots.

\subsection{DoF Definition}
Making use of the same notation as in \cite{Inf_Theo} and \cite{network_info}, a rate pair $(R_1{,}R_2)$ is said to be achievable in an $L$-subband BC with arbitrary imperfect CSIT qualities if there exists a code sequence $(2^{nR_1}{,}2^{nR_2}{,}n)$ such that
\begin{itemize}
\item \emph{Codebook construction:} There is one message set for each user. To be specific, $M_1$ for user 1 is uniformly distributed in the set $\mathcal{M}_1{\triangleq}[1:2^{nR_1}]$ and $M_2$, intended for user 2, is similarly distributed in the set $\mathcal{M}_2{\triangleq}[1:2^{nR_2}]$.

\item \emph{Encoding:} The encoder randomly chooses a message $M_1$ from $\mathcal{M}_1$ and generate $u_1^{n}(M_1,\hat{\mathbf{\mathcal{S}}}_1^n)$ according to the probability $\Pi_{i{=}1}^{n}p(u_i|\hat{\mathbf{\mathcal{S}}}_i)$. At the same time $v_1^{n}(M_2,\hat{\mathbf{\mathcal{S}}}_1^n)$ with the probability $\Pi_{i{=}1}^{n}p(v_i|\hat{\mathbf{\mathcal{S}}}_i)$ is generated as a function of $M_2$ which is randomly chosen from $\mathcal{M}_2$. Finally, the codeword $x_1^{n}(u_1^{n}{,}v_1^{n}{,}\hat{\mathbf{\mathcal{S}}}_1^n)$ is generated with the probability $\Pi_{i{=}1}^{n}p(x_i|u_i,v_i,\hat{\mathbf{\mathcal{S}}}_1^n)$.

\item \emph{Decoding:} Receiver 1 wishes to decode $M_1$ and declares message $\hat{M}_1(y_1^n{,}\hat{\mathbf{\mathcal{S}}}_1^n{,}\tilde{\mathbf{\mathcal{S}}}_1^n)$ is sent if it is the unique message such that $y_1^n$ and $u_1^n(\hat{M}_1{,}\hat{\mathbf{\mathcal{S}}}_1^n)$ are jointly typical. Similarly, receiver 2 declares message $\hat{M}_2(z_1^n{,}\hat{\mathbf{\mathcal{S}}}_1^n{,}\tilde{\mathbf{\mathcal{S}}}_1^n)$ is sent if it is the unique message such that $z_1^n$ and $v_1^n(\hat{M}_2{,}\hat{\mathbf{\mathcal{S}}}_1^n)$ are jointly typical. Otherwise, an error $P_e^{(n)}$ will occur. By the Law of Large Numbers, we have $P_e^{(n)}{\to}0$ when $n{\to}\infty$.
\end{itemize}

The capacity region, $\mathcal{C}$, is formed by all the achievable rate pairs. The \emph{DoF} region, $\mathcal{D}$, is accordingly defined on a per-channel-use basis as follows
\begin{equation}
\mathcal{D}\triangleq\left\{(d_1{,}d_2)|{\forall}(w_1{,}w_2){\in}\mathbb{N}^2{,}{\forall}(R_1{,}R_2){\in}\mathcal{C}{,}
w_1d_1{+}w_2d_2{\leq}\lim_{P\to\infty}\sup\frac{w_1R_1{+}w_2R_2}{r\log P}\right\}, \label{eq:dof_def}
\end{equation}
where $r$ is the channel uses actually employed to achieve the rate pair $(R_1{,}R_2)$.

\subsection{Problem Model}\label{P_model}
The average CSIT quality of user 1 and user 2 are respectively expressed as $a_e{=}\frac{1}{L}\sum_{j{=}1}^La_j$ and $b_e{=}\frac{1}{L}\sum_{j{=}1}^Lb_j$. Without the loss of generality, in the rest of this paper, we consider $a_j{\geq}b_j$ in subband 1 to $l$ ($l{\leq}L$) and $a_j{\leq}b_j$ for the remaining subbands. For convenience, we denote $q_j^+{\triangleq}a_j{-}b_j$ if $a_j{\geq}b_j$ (namely for $j{\leq}l$) while $q_j^-{\triangleq}b_j{-}a_j$ if $a_j{\leq}b_j$ (namely for $l{+}1{\leq}j{\leq}L$). Then, we have $\{q^+\}{\triangleq}\{q_1^+{,}q_2^+{,}\cdots{,}q_l^+\}$ and $\{q^-\}{\triangleq}\{q_{l{+}1}^-{,}q_{l{+}2}^-{,}\cdots{,}q_{L}^-\}$.

For any positive integer $L$, we define two classes of problems as follows
\newtheorem{mydef}{Definition}
\begin{mydef}\label{PLpro}
\emph{$\mathcal{P}_L$ Problem}: Achieve the optimal \emph{DoF} region in a scenario such that $a_e{=}b_e$.
\end{mydef}
\begin{mydef}\label{QLpro}
\emph{$\mathcal{Q}_L$ Problem}: Achieve the optimal \emph{DoF} region in a scenario such that $a_e{\neq}b_e$. Note that for $a_e{>}b_e$ (resp. $a_e{<}b_e$), it is called $\mathcal{Q}_L^+$ (resp. $\mathcal{Q}_L^-$) Problem hereafter.
\end{mydef}

A $\mathcal{P}_L$ problem considers the general $L$-subband scenario with $a_e{=}b_e$. Specifically, if there exists a subset of the subbands, denoted as $\mathcal{J}$, such that $\sum_{j_1{\in}\mathcal{J}}a_{j_1}{=}\sum_{j_1{\in}\mathcal{J}}b_{j_1}$, the $\mathcal{P}_L$ problem can be solved as a combination of a $\mathcal{P}_{|\mathcal{J}|}$ and a $\mathcal{P}_{L{-}|\mathcal{J}|}$ problems. If no such subset $\mathcal{J}$ is found, the $\mathcal{P}_L$ problem considers the most complicated $L$-subband scenario with $a_e{=}b_e$. For instance, when $L{=}2$ and $a_1{+}a_2{=}b_1{+}b_2$, there generally exist two possible CSIT patterns: 1) $a_1{=}b_1$ and $a_2{=}b_2$; 2) $a_1{\neq}b_1$ and $a_2{\neq}b_2$. The first case refers to two $\mathcal{P}_1$ problems and the optimal scheme is obtained by performing the solution to $\mathcal{P}_1$ problem twice (separately and independently in subband 1 and 2). However, for the second case, the transmitted signal in each subband is correlated to each other (see Section \ref{P2}). Similarly, a $\mathcal{Q}_L^+$ problem considers the general $L$-subband scenario with $a_e{>}b_e$. In other words, for a $\mathcal{P}_L$ and a $\mathcal{Q}_L^+$ problem, the transmitted signals vary according the actual CSIT quality pattern (formed by the frequency-user grid as shown in Figure \ref{fig:system}). More details of the achievabilities in a $\mathcal{P}_L$ problem and a $\mathcal{Q}_L^+$ are shown in Section \ref{achp} and \ref{achq} respectively.

\subsection{Main Results}
\newtheorem{mytheorem}{Theorem}
\begin{mytheorem} \label{theo1}
The optimal \emph{DoF} region, $\mathcal{D}$, in a $L$-subband frequency correlated BC with imperfect varying CSIT is specified by
\begin{align}
d_1{+}d_2{\leq}&1{+}\frac{1}{L}\min(\sum_{j{=}1}^La_j{,}\sum_{j{=}1}^Lb_j),\label{eq:theo1}\\
d_1{\leq}&1,\label{eq:single1}\\
d_2{\leq}&1,\label{eq:single2}
\end{align}
where $a_j$ and $b_j$ are respectively the quality of the CSIT of user 1 and user 2 in subband $j$ and $L$ can be any integer values.
\end{mytheorem}

Note that the optimal \emph{DoF} region is bounded by the minimum average CSIT quality between user 1 and user 2. This result gives an affirmative answer to the conjecture in \cite{Tandon12} that the sum \emph{DoF} is $1$ in a two-user MISO BC with fixed $PN$ CSIT state. The converse is provided in Section \ref{conv}. The achievability is discussed in Section \ref{achp} and \ref{achq}, for $\mathcal{P}_L$ ($a_e{=}b_e$) and $\mathcal{Q}_L$ ($a_e{\neq}b_e$) problem respectively. The following corollary provides an instructive insight into the formation of the optimal \emph{DoF} region.
\newtheorem{mycorollary}{Corollary}
\begin{mycorollary} \label{corollary_wsum}
The optimal \emph{DoF} region in the frequency correlated BC with imperfect CSIT can be interpreted as a weighted sum of three basis optimal \emph{DoF} regions
\begin{equation}
\mathcal{D}{=}\frac{1}{L}(\bar{r}\bar{\mathcal{D}}{+}\hat{r}\hat{\mathcal{D}}{+}\tilde{r}\tilde{\mathcal{D}}),\label{eq:corollary}
\end{equation}
where $\bar{\mathcal{D}}$, $\hat{\mathcal{D}}$ and $\tilde{\mathcal{D}}$ refer to the optimal \emph{DoF} region for a CSIT state of $PP$, $PN/NP$ and $NN$ respectively and $\bar{r}$, $\hat{r}$ and $\tilde{r}$ are the corresponding weights, given as
\begin{align}
\bar{\mathcal{D}}{:}&d_1{\leq}1{,}d_2{\leq}1,\\
\hat{\mathcal{D}}{:}&d_1{+}d_2{\leq}\frac{3}{2}{,}d_1{\leq}1{,}d_2{\leq}1,\\
\tilde{\mathcal{D}}{:}&d_1{+}d_2{\leq}1,\\
\bar{r}{=}&\sum_{j{=}1}^L\min(a_j{,}b_j),\label{eq:rbar}\\
\hat{r}{=}&2\min(\sum_{j{=}1}^lq_{j}^+{,}\sum_{j{=}l{+}1}^Lq_{j}^-),\label{eq:rhat}\\
\tilde{r}{=}&L{-}\bar{r}{-}\hat{r}.\label{eq:rtilde}
\end{align}

\end{mycorollary} 

\section{Converse of Theorem \ref{theo1}}\label{conv}
The objective of this section is to provide the converse of Theorem \ref{theo1}. Before going into the details, we highlight the key ingredients in the derivation as
\begin{itemize}
\item Nair-Gamal bound \cite{NairGamal}: provides an upper-bound to the \emph{DoF} region in a general BC;
\item Extremal Inequality: maximizes a weighted difference of two different entropies;
\item Lemma 1 in \cite{Ges12}: upper- and lower-bound the entropy.
\end{itemize}

Let us revisit the converse in previous works. In \cite{HJSV09}, the \emph{DoF} region in the BC without CSIT is upper-bounded by considering one user's observation is degraded compared to the other's. In the BC with delayed CSIT \cite{Tse10}\cite{Ges12}\cite{Gou12}, the outer-bound is obtained through the genie-aided model where one user's observation is provided to the other, thus establishing a physically degraded BC to remove the delayed CSIT.

However, in this contribution, those methods are not adopted since the transmitter does not have delayed CSIT and the BC with imperfect CSIT cannot be simply considered as a degraded BC. Instead, we follow the assumption in \cite{Lapidoth}: We first consider that user 2 knows the message intended for user 1, which leads to an outer-bound denoted by $\mathbb{D}_1$; Then by assuming that user 1 knows user 2's desired message, we can have another region $\mathbb{D}_2$. The final \emph{DoF} outer-bound results from the intersection of them, i.e. $\mathbb{D}{=}\mathbb{D}_1{\Cap}\mathbb{D}_2$. This assumption is consistent with the derivation in Korner-Marton bound (Theorem 5 and Appendix I in \cite{Korner_Marton}) and Nair-Gamal bound (Theorem 2.1 and 3.1 in \cite{NairGamal}, proof given in the Appendix of Lecture Notes 9 in \cite{network_info}). Both of these two bounds provide an outer-bound to the general discrete memoryless broadcast channel and Nair-Gamal bound is said to be in general contained in Korner-Marton bound \cite{NairGamal}. As a consequence, we aim at finding the following bounds
\begin{align}
R_1{+}R_2&\leq I(U;Y|V){+}I(V;Z),\label{eq:generalRs1}\\
R_1{+}R_2&\leq I(U;Y){+}I(V;Z|U).\label{eq:generalRs2}
\end{align}
The key challenge lies in finding the auxiliary variables $U$ and $V$.

Assuming user 2 has the knowledge of $M_1$ and according to Fano's Inequality, we have
\begin{align}
nR_1&{\leq}I(M_1;Y_1^n|\hat{\mathcal{S}}_1^n{,}\tilde{\mathcal{H}}_1^n)\\
&{=}I(M_1;Y_1^n|\hat{\mathcal{S}}_1^n{,}\tilde{\mathcal{H}}_1^n{,}\tilde{\mathcal{G}}_1^n)\label{eq:r1}\\
nR_2&{\leq}I(M_2;Z_1^n|M_1{,}\hat{\mathcal{S}}_1^n{,}\tilde{\mathcal{G}}_1^n)\\
&{=}I(M_2;Z_1^n|M_1{,}\hat{\mathcal{S}}_1^n{,}\tilde{\mathcal{G}}_1^n{,}\tilde{\mathcal{H}}_1^n)\label{eq:r2}\\
n(R_1{+}R_2)&{\leq}I(M_1;Y_1^n|\hat{\mathcal{S}}_1^n{,}\tilde{\mathcal{H}}_1^n{,}\tilde{\mathcal{G}}_1^n){+}
I(M_2;Z_1^n|M_1{,}\hat{\mathcal{S}}_1^n{,}\tilde{\mathcal{G}}_1^n{,}\tilde{\mathcal{H}}_1^n)\\
&{=}I(M_1;Y_1^n|\mathcal{S}){+}I(M_2;Z_1^n|M_1{,}\mathcal{S}),\label{eq:rs0}
\end{align}
where \eqref{eq:r1} follows the fact that $Y_1^n{\to}\{\hat{\mathcal{S}}_1^n{,}\tilde{\mathcal{H}}_1^n\}{\to}\tilde{\mathcal{G}}_1^n$ forms a Markov chain such that $Y_1^n$ is independent of $\tilde{\mathcal{G}}_1^n$ conditioned on $\{\hat{\mathcal{S}}_1^n{,}\tilde{\mathcal{H}}_1^n\}$. \eqref{eq:r2} follows similarly. In \eqref{eq:rs0}, $\{\hat{\mathcal{S}}_1^n{,}\tilde{\mathcal{G}}_1^n{,}\tilde{\mathcal{H}}_1^n\}$ is replaced by $\mathcal{S}$. \eqref{eq:rs0} is bounded by
\begin{align}
n(R_1{+}R_2)&{\leq}\sum_{j{=}1}^n\{I(M_1{,}Y_1^{j{-}1}{,}Z_{j{+}1}^n{,}\hat{\mathcal{S}}_1^n;Y_j|\mathcal{S}_1^n){+}
I(M_2{,}Y_1^{j{-}1}{,}Z_{j{+}1}^n{,}\hat{\mathcal{S}}_1^n;Z_j|M_1{,}\mathcal{S}_1^n{,}Y_1^{j{-}1}{,}Z_{j{+}1}^n)\}\label{eq:appendix}\\
&{=}\sum_{j{=}1}^n\{I(U_j;Y_j|\mathcal{S}_1^n){+}I(V_j;Z_j|U_j{,}\mathcal{S}_1^n)\}.\label{eq:gamalexpress}
\end{align}
The derivation of \eqref{eq:appendix} is provided in the Appendix. Now, we have found the auxiliary variables as
\begin{align}
U_j{\triangleq}\{M_1{,}Y_1^{j{-}1}{,}Z_{j{+}1}^n{,}\hat{\mathcal{S}}_1^n\},\\
V_j{\triangleq}\{M_2{,}Y_1^{j{-}1}{,}Z_{j{+}1}^n{,}\hat{\mathcal{S}}_1^n\}.
\end{align}
Continuing deriving \eqref{eq:gamalexpress}, we have
\begin{align}
n(R_1{+}R_2){\leq}&\sum_{j{=}1}^n\underbrace{h(Y_j|\mathcal{S}_1^n)}_{{\leq}{\log}P}{-}h(Y_j|\mathcal{S}_1^n{,}U_j){+}
h(Z_j|\mathcal{S}_1^n{,}U_j){-}\underbrace{h(Z_j|\mathcal{S}_1^n{,}U_j{,}V_j)}_{{\leq}o({\log}P)}\\
{\leq}&n{\log}P{+}\sum_{j{=}1}^n\{h(Z_j|\mathcal{S}_1^n{,}U_j){-}h(Y_j|\mathcal{S}_1^n{,}U_j)\}.\label{eq:rs1}
\end{align}
Next, we focus on the terms in the summation of \eqref{eq:rs1} and upper-bound them using a similar derivation as in \cite{Ges12}. For convenience, we give up the index $j$. Consequently,
\begin{align}
h(Z|\mathcal{S}{,}U){-}h(Y|\mathcal{S}{,}U)
{\leq}&\max_{P_UP_{\mathbf{x}|U}}\{h(Z|U{,}\mathcal{S}){-}h(Y|U{,}\mathcal{S})\}\\
{\leq}&\max_{P_U}\mathcal{E}_U\{\max_{P_{\mathbf{x}|U}}h(Z|U{=}U^*{,}\mathcal{S}){-}h(Y|U{=}U^*{,}\mathcal{S})\}\\
{\leq}&\max_{P_U}\mathcal{E}_U\{\max_{P_{\mathbf{x}|U}}\mathcal{E}_{\mathcal{S}|U}
[h(Z|U{=}U^*{,}\mathcal{S}{=}\mathcal{S}^*){-}h(Y|U{=}U^*{,}\mathcal{S}{=}\mathcal{S}^*)]\}\\
{=}&\max_{P_U}\mathcal{E}_U\{\max_{P_{\mathbf{x}|U}}\mathcal{E}_{\mathcal{S}|\hat{\mathcal{S}}}
[h(\mathbf{g}^H\mathbf{x}{+}\epsilon_2|U{=}U^*){-}h(\mathbf{h}^H\mathbf{x}{+}\epsilon_1|U{=}U^*)]\}\\
{\leq}&\max_{P_U}\mathcal{E}_U\{\max_{\mathbf{0}{\preceq}\mathbf{C}{,}tr(\mathbf{C}){\leq}P}
\max_{\stackrel{P_{\mathbf{x}|U}}{Cov(\mathbf{x}|U)\preceq\mathbf{C}}}\mathcal{E}_{\mathcal{S}|\hat{\mathcal{S}}}
[h(\mathbf{g}^H\mathbf{x}{+}\epsilon_2|U{=}U^*){-}h(\mathbf{h}^H\mathbf{x}{+}\epsilon_1|U{=}U^*)]\}\\
{\leq}&\max_{P_U}\mathcal{E}_U\{\max_{\mathbf{0}{\preceq}\mathbf{C}{,}tr(\mathbf{C}){\leq}P}\mathcal{E}_{\mathcal{S}|\hat{\mathcal{S}}}
[{\log}(1{+}\mathbf{g}^H\mathbf{K}\mathbf{g}){-}{\log}(1{+}\mathbf{h}^H\mathbf{K}\mathbf{h})]\},\label{eq:extremal}\\
{\leq}&\mathcal{E}_{\hat{\mathcal{S}}}\{\max_{0{{\preceq}\mathbf{K}{,}tr(\mathbf{K}){\leq}P}}\mathcal{E}_{\mathcal{S}|\hat{\mathcal{S}}}
[{\log}(1{+}\mathbf{g}^H\mathbf{K}\mathbf{g}){-}{\log}(1{+}\mathbf{h}^H\mathbf{K}\mathbf{h})]\},\label{eq:extremal2}
\end{align}
where $\mathbf{K}$ is the covariance matrix of $\mathbf{x}$ (i.e. $Cov(\mathbf{x}|U){=}\mathbf{K}$) and $\mathbf{C}$ is a semi-definite matrix, which is regarded as the constraint of $\mathbf{K}$. \eqref{eq:extremal} is derived according to the fact 1) $\mathbf{x}{\to}U{\to}\mathcal{S}$ forms a Markov chain so that $\mathcal{S}$ is independent of $\mathbf{x}$ conditioned on $U$; 2) With a constrained covariance, a Gaussian distributed $\mathbf{x}$ conditioned on $U$ is the optimal solution to the maximization of the weighted difference in \eqref{eq:extremal} for any positive semi-definite $\mathbf{C}$, based on the proof of Corollary 6 in \cite{Extremal}.

Using Lemma 1 in \cite{Ges12}, we can respectively upper- and lower-bound the first and second terms in \eqref{eq:extremal2} as
\begin{align}
\mathcal{E}_{\mathcal{S}|\hat{\mathcal{S}}}{\log}(1{+}\mathbf{g}^H\mathbf{K}\mathbf{g}){\leq}&
{\log}(1{+}\lambda_1\mathcal{E}[||\hat{\mathbf{g}}||^2]){+}o(1),\label{eq:upper}\\
\mathcal{E}_{\mathcal{S}|\hat{\mathcal{S}}}{\log}(1{+}\mathbf{h}^H\mathbf{K}\mathbf{h}){\geq}&
{\log}(1{+}e^{-\gamma}\lambda_1\mathcal{E}[||\tilde{\mathbf{h}}||^2]){+}o(1),\label{eq:lower}
\end{align}
where $\gamma$ is a constant, $\lambda_1$ is the largest eigen-value of the covariance matrix $\mathbf{K}$. Substituting the terms in \eqref{eq:extremal2} with \eqref{eq:upper} and \eqref{eq:lower}, we can have the $j$th term in the summation of \eqref{eq:rs1} upper-bounded by
\begin{align}
h(Z_j|\mathcal{S}_1^n{,}U_j){-}h(Y_j|\mathcal{S}_1^n{,}U_j){\leq}&
{\log}\frac{1{+}\lambda_1\mathcal{E}[||\hat{\mathbf{g}}_j||^2]}{1{+}e^{-\gamma}\lambda_1\mathcal{E}[||\tilde{\mathbf{h}}_j||^2]}\\
{\approx}&a_j{\log}P.\label{eq:aj}
\end{align}
Applying \eqref{eq:aj} to all the terms in \eqref{eq:rs1}, the sum rate is upper-bounded by
\begin{align}
n(R_1{+}R_2){\leq}&n{\log}P{+}\sum_{j{=}1}^na_j{\log}P\\
R_1{+}R_2{\leq}&{\log}P{+}\frac{1}{n}\sum_{j{=}1}^na_j{\log}P.\label{eq:1overL}
\end{align}
When $n$ tends to infinity, the $L$-subband scenario defined in Figure \ref{fig:system} repeats infinite times. Consequently, the CSIT state in each subband appears $n\times\frac{1}{L}$ times and \eqref{eq:1overL} can be rewritten as
\begin{equation}
R_1{+}R_2{\leq}{\log}P{+}\frac{1}{n}\sum_{j{=}1}^L\frac{n}{L}a_j{\log}P{=}{\log}P{+}\frac{1}{L}\sum_{j{=}1}^La_j{\log}P,
\end{equation}
Accordingly, the \emph{DoF} region is specified as follows
\begin{equation}
\mathbb{D}_1:d_1{+}d_2{\leq}1{+}a_e{=}1{+}\frac{\sum_{j{=}1}^La_j}{L}.\label{eq:D1}
\end{equation}

Switching the role of each user results in the sum rate and \emph{DoF} region specified as
\begin{align}
n(R_1{+}R_2){\leq}&\sum_{j{=}1}^n\{I(U_j;Y_j|\mathcal{S}_1^n{,}V_j){+}I(V_j;Z_j|\mathcal{S}_1^n)\}\\
{\leq}&n{\log}P{+}\sum_{j{=}1}^nh(Y_j|\mathcal{S}_1^n{,}V_j){-}h(Z_j|\mathcal{S}_1^n{,}V_j)\\
{\leq}&n{\log}P{+}\sum_{j{=}1}^nb_j{\log}P,\label{eq:Rs2}\\
R_1{+}R_2{\leq}&{\log}P{+}\frac{1}{L}\sum_{j{=}1}^Lb_j{\log}P,\\
\mathbb{D}_2:d_1{+}d_2{\leq}&1{+}b_e{=}1{+}\frac{\sum_{j{=}1}^Lb_j}{L}.\label{eq:D2}
\end{align}

Taking the intersection of $\mathbb{D}_1$ and $\mathbb{D}_2$ results in \eqref{eq:theo1}. Together with the single-user constraint, Theorem \ref{theo1} holds.

$\hfill\Box$ 

\section{A Weighted-Sum Interpretation of the Optimal \emph{DoF} Region}\label{wsum}
In this section, we provide an insight into the formation of the optimal \emph{DoF} region. According to the particular CSIT setting, each subband is considered as composed of three parallel subchannels with different fraction of channel use. The \emph{DoF} region of each subchannel has been found in previous work. We will show that the optimal \emph{DoF} region stated in Theorem \ref{theo1} can be calculated as a weighted sum of the \emph{DoF} region of each subchannel.

\subsection{Intuition: Channel Decomposition}
\begin{figure}[t]
\renewcommand{\captionfont}{\small}
\centering
\includegraphics[height=5cm,width=10cm]{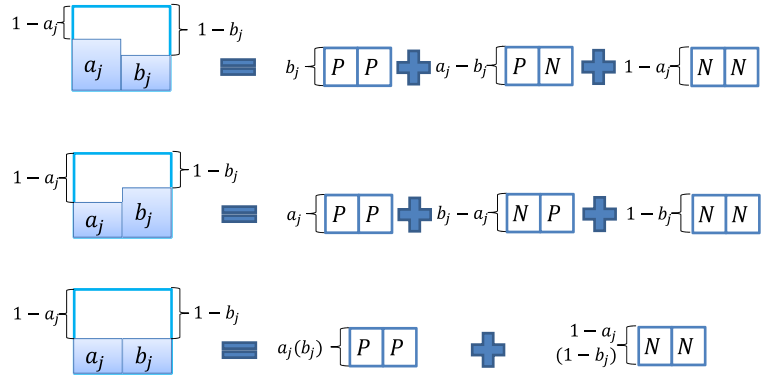}
\caption{Subband $j$ is decomposed to subchannels with states $PP$, $NP/PN$ and $NN$, each with an amount of channel use determined according to the CSIT qualities.}\label{fig:decomp}
\end{figure}
In this part, we decompose the channel in each subband following the intuition that the imperfect CSIT with error variance $P^{-\alpha}$ can be considered as perfect for $\alpha$ ($0{\leq}\alpha{\leq}1$) channel use (i.e. the transmit power is reduced to $\mathcal{E}[||\mathbf{s}||^2]{\leq}P^\alpha$). We can see this by simply sending one private message per user using ZFBF precoding and with power $P^{\alpha}$. Since $\mathcal{E}[|\mathbf{h}_j^H\hat{\mathbf{h}}_j^\bot|]{\sim}P^{-\alpha}$ and $\mathcal{E}[|\mathbf{g}_j^H\hat{\mathbf{g}}_j^\bot|]{\sim}P^{-\alpha}$, both users can recover their private messages subject to noise. Therefore, the rate $\alpha{\log}P$ is achieved per user. As only $\alpha$ channel has been used, full \emph{DoF} region (i.e. $d_1{=}1$ and $d_2{=}1$) is obtained according to \eqref{eq:dof_def}. This is in fact a generalization of the fact that  full \emph{DoF} region can be obtained if the variance of CSIT error is scaled as $SNR^{-1}$ \cite{Ges12}.

Consequently, a subband $j$ with the CSIT error scaling as $P^{-a_j}$ and $P^{-b_j}$ for user 1 and 2 respectively, is decomposed as shown in Figure \ref{fig:decomp}. The transmitter is assumed to have perfect knowledge of the CSI of user 1 for $a_j$ channel use while for the remaining $1{-}a_j$ channel use, no CSIT of user 1 is available. The same approach is employed for user 2. It results three subchannels, each of which can be interpreted using the same notation as in \cite{Tandon12} ($NN$, $PN/NP$ and $PP$).
\begin{itemize}
  \item $\tilde{j}$: $NN$ channel, no CSIT of either user, with channel use $1{-}\max(a_j{,}b_j)$;
  \item $\hat{j}$: $PN$ (resp. $NP$) channel, perfect CSIT of user 1 (resp. 2) but no CSIT of use 2 (resp. 1), with channel use $q_j^+$ (resp. $q_j^-$);
  \item $\bar{j}$: $PP$ channel, perfect CSIT of both users, with channel use $\min(a_j{,}b_j)$.
\end{itemize}

In this way, the original $L$-subband scenario becomes the product of those parallel subchannels. The \emph{DoF} region is obtained as the weighted-sum of that in each subchannel.

\subsection{\emph{DoF} Regions of the Subchannels}\label{dof_subchannel}
We split the rate of each user into three parts, namely $R_1{=}\tilde{R}_1{+}\hat{R}_1{+}\bar{R}_1$ and $R_2{=}\tilde{R}_2{+}\hat{R}_2{+}\bar{R}_2$, where $(\tilde{R}_1{,}\tilde{R}_2)$ represents the rate pair achieved in the subchannel with state $NN$, $(\hat{R}_1{,}\hat{R}_2)$ refers to the rate pair achieved in the subchannel with alternating $PN/NP$ state while $(\bar{R}_1{,}\bar{R}_2)$ is the rate pair achieved in the subchannel with state $PP$. The message intended to user 1 and 2 is therefore generated from a set jointly formed by $\tilde{\mathcal{M}}_1{\times}\hat{\mathcal{M}}_1{\times}\bar{\mathcal{M}}_1{=}[1{:}2^{n\tilde{R}_1}][1{:}2^{n\hat{R}_1}][1{:}2^{n\bar{R}_1}]$ and $\tilde{\mathcal{M}}_2{\times}\hat{\mathcal{M}}_2{\times}\bar{\mathcal{M}}_2{=}[1{:}2^{n\tilde{R}_2}][1{:}2^{n\hat{R}_2}][1{:}2^{n\bar{R}_2}]$ respectively. The subsets ($\tilde{\mathcal{M}}_k$, $\hat{\mathcal{M}}_k$ and $\bar{\mathcal{M}}_k$) are independent of each other for $k{=}1{,}2$. $(\tilde{R}_1{,}\tilde{R}_2)$, $(\hat{R}_1{,}\hat{R}_2)$ and $(\bar{R}_1{,}\bar{R}_2)$ respectively result in the \emph{DoF} region $\tilde{\mathcal{D}}$, $\hat{\mathcal{D}}$ and $\bar{\mathcal{D}}$.

\subsubsection{Subchannel $\bar{j}$}
When the transmitter has perfect CSI of both users, the optimal \emph{DoF} region is expressed as follows
\begin{align}
\bar{\mathcal{D}}:& d_1{\leq}1{,}d_2{\leq}1,\label{eq:DPP}
\end{align}
which can be achieved via ZFBF. The total amount of channel use of the subchannels with $PP$ state is
\begin{equation}
\bar{r}{=}\sum_{j{=}1}^L\min(a_j{,}b_j).\label{eq:rPP}
\end{equation}

\subsubsection{Subchannel $\hat{j}$}
\begin{figure}[t]
\renewcommand{\captionfont}{\small}
\centering
\includegraphics[height=4cm,width=15cm]{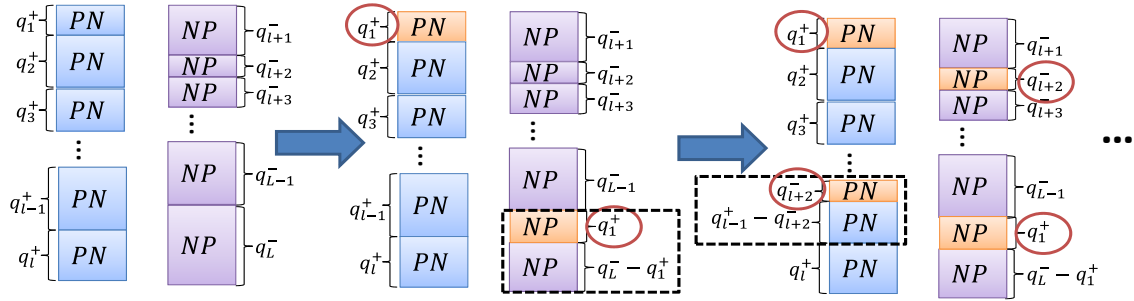}
\caption{An example of further decomposing subchannel $\hat{j}$ to have multiple equivalent alternating $PN/NP$ scenario.}\label{fig:further_decomp}
\end{figure}
In this class of subchannels, the transmitter has perfect knowledge of the CSI of user 1 or (exclusive) user 2. As a reminder, we assume $a_j{\geq}b_j$ in subband 1 to $l$ ($l{\leq}L$) and $a_j{\leq}b_j$ for the remaining subbands. Following the way where channels are decomposed (as in Figure \ref{fig:decomp}), there are in total $l$ $PN$ subchannels, each with channel use $q_j^+{=}a_j{-}b_j{,}j{\leq}l$ and $L{-}l$ $NP$ subchannels, each with channel use $q_j^-{=}b_j{-}a_j{,}l{+}1{\leq}j{\leq}L$. Literature \cite{Tandon12} provides an optimal \emph{DoF} region of the alternating $PN/NP$ scenario, which can be achieved by the simple $S_3^{3/2}$ scheme. This bound is denoted by $\hat{\mathcal{D}}$ and expressed as
\begin{align}
\hat{\mathcal{D}}:& d_1+d_2\leq\frac{3}{2},d_1\leq 1,d_2\leq1.\label{eq:DPNNP}
\end{align}
However, this region is optimal for the alternating $PN/NP$ scenario where each $PN$ and $NP$ subchannel have the same amount of channel use, namely ${\forall}j_1{\in}[1{,}l]{,}{\exists}j_2{\in}[l{+}1{,}L]$ such that $q_{j_1}^+{=}q_{j_2}^-$. In a general $L$-subband scenario (Definition \ref{PLpro} and \ref{QLpro}), this condition does not necessarily hold. Hence, we aim at showing that \eqref{eq:DPNNP} still optimally bounds the \emph{DoF} region of the $PN$ and $NP$ subchannels. To that end, we further decompose the subchannels in order to find the alternating $PN/NP$ scenario.

Figure \ref{fig:further_decomp} shows an example of the further decomposition. Firstly, subchannel $\hat{L}$ is decomposed into two $NP$ subchannels, namely $\hat{L}(1)$ and $\hat{L}(2)$, respectively with fraction of channel use $q_1^+$ and $q_L^-{-}q_1^+$. In this way, subchannel $\hat{1}$ and $\hat{L}(1)$ form an alternating $PN/NP$ scenario where $PN$ and $NP$ states have equal amount of channel use. Secondly, subchannel $\hat{l{-}1}$ is decomposed into two $PN$ subchannels, namely $\hat{(l{-}1)}(1)$ and $\hat{(l{-}1)}(2)$, respectively with fraction of channel use $q_{l{+}2}^-$ and $q_{l{-}1}^+{-}q_{l{+}2}^-$. Then we consider subchannel $\hat{(l{-}1)}(1)$ and $\hat{l{+}2}$ as an alternating $PN/NP$ scenario. Such process can be repeated till no $PN$ subchannels or $NP$ subchannels remains. Consequently, multiple alternating $PN/NP$ scenario are found, with the total amount of channel use (denoted as $\hat{r}$)
\begin{equation}
\hat{r}{=}2\min(\sum_{j{=}1}^lq_j^+{,}\sum_{j{=}l{+}1}^Lq_j^-).\label{eq:rPNNP}
\end{equation}

When $\sum_{j{=}1}^lq_{j}^+{\neq}\sum_{j{=}l{+}1}^Lq_{j}^-$, for instance a $\mathcal{Q}_L$ problem, the remaining $\hat{r}^\prime{=}|\sum_{j{=}1}^lq_{j}^+{-}\sum_{j{=}l{+}1}^Lq_{j}^-|$ channel use of $PN$ (or $NP$) subchannels are merged with the subchannels with state $NN$. This is because the \emph{DoF} region of a $PN$ (or $NP$) subchannel is identical to that of a $NN$ subchannel, according to Theorem \ref{theo1} applied to the case where $a_{1{:}L}{=}1$ and $b_{1{:}L}{=}0$.

\subsubsection{Subchannel $\tilde{j}$}
In subchannel $\tilde{j}$, the transmitter has no knowledge of the CSI of both users. The optimal \emph{DoF} region (denoted as $\tilde{\mathcal{D}}$) has been studied in \cite{VV11a} and \cite{HJSV09}, which can be achieved by simply performing FDMA. This optimal \emph{DoF} region writes as
\begin{equation}
\tilde{\mathcal{D}}: d_1+d_2\leq1. \label{eq:DNN}
\end{equation}
Subband $\tilde{j}$ have $L{-}\sum_{j{=}1}^L\max(a_j{,}b_j)$ channel use in total. Combining with the $\hat{r}^\prime$ channel use of $PN$ (or $NP$) subchannels, the total amount of channel use where $\tilde{\mathcal{D}}$ is optimal is
\begin{equation}
\tilde{r}{=}L{-}\sum_{j{=}1}^L\max(a_j{,}b_j){+}\hat{r}^\prime.\label{eq:rNN}
\end{equation}

\subsection{Weighted-Sum}
As the rate per user can be expressed as $R_k{=}\frac{1}{L}(\bar{r}\bar{R}_k{+}\hat{r}\hat{R}_k{+}\tilde{r}\tilde{R}_k)$, we interpret the optimal \emph{DoF} region as
\begin{align}
\mathcal{D}{=}&\frac{1}{L}(\bar{r}\bar{\mathcal{D}}{+}\hat{r}\hat{\mathcal{D}}{+}\tilde{r}\tilde{\mathcal{D}})\nonumber\\
{=}&\frac{\sum_{j{=}1}^L\min(a_j{,}b_j)}{L}\bar{\mathcal{D}}{+}\frac{2{\min}(\sum_{j{=}l{+}1}^Lq_{j}^-{,}\sum_{j{=}1}^lq_{j}^+)}{L}\hat{\mathcal{D}}{+}
\frac{L{-}\sum_{j{=}1}^L\max(a_j{,}b_j){+}|\sum_{j{=}1}^lq_{j}^+{-}\sum_{j{=}l{+}1}^Lq_{j}^-|}{L}\tilde{\mathcal{D}},\label{eq:wsum}
\end{align}
\begin{figure}[t]
\renewcommand{\captionfont}{\small}
\centering
\includegraphics[height=6.5cm,width=8cm]{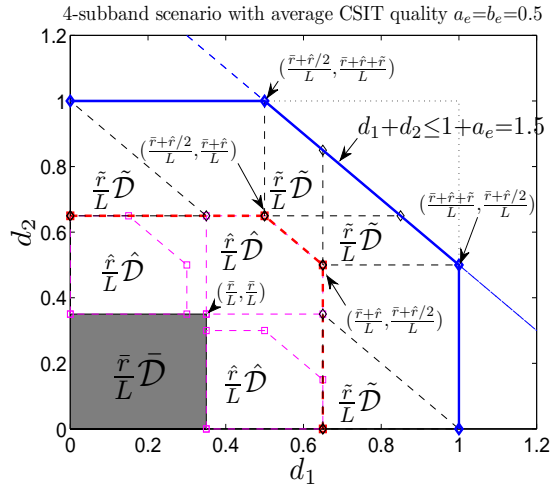}
\caption{The composition of the optimal \emph{DoF} of a 4-subband scenario, with $(a_1{,}b_1){=}(0.7{,}0.3)$, $(a_2{,}b_2){=}(0.6{,}0.4)$, $(a_3{,}b_3){=}(0.4{,}0.7)$ and $(a_4{,}b_4){=}(0.3{,}0.6)$, thus $(a_e{,}b_e){=}(0.5{,}0.5)$.}\label{fig:D_form1}
\end{figure}

Figure \ref{fig:D_form1} illustrates the composition of $\mathcal{D}$. The grey square area depicts the region $\frac{\bar{r}}{L}\bar{\mathcal{D}}$, specified by the corner point $(\frac{\bar{r}}{L}{,}\frac{\bar{r}}{L})$. All the valid points inside $\frac{\bar{r}}{L}\bar{\mathcal{D}}$ are expanded to a magenta polygon representing $\frac{\hat{r}}{L}\hat{\mathcal{D}}$. This expansion results in the bound shown in the dashed red curve with square points, outlined by the corner points $(\frac{\bar{r}{+}\hat{r}/2}{L}{,}\frac{\bar{r}{+}\hat{r}}{L})$ and $(\frac{\bar{r}{+}\hat{r}}{L}{,}\frac{\bar{r}{+}\hat{r}/2}{L})$. Then, every point on this bound is further expanded to a black triangle area referring to the \emph{DoF} region $\frac{\tilde{r}}{L}\tilde{\mathcal{D}}$. Outlining all the expanded area, we can obtain $\mathcal{D}$ specified by the solid blue curve with diamond points $(\frac{\bar{r}{+}\hat{r}/2}{L}{,}\frac{\bar{r}{+}\hat{r}{+}\tilde{r}}{L})$ and $(\frac{\bar{r}{+}\hat{r}{+}\tilde{r}}{L}{,}\frac{\bar{r}{+}\hat{r}/2}{L})$. Replacing $\bar{r}$, $\hat{r}$ and $\tilde{r}$ with \eqref{eq:rPP}, \eqref{eq:rPNNP} and \eqref{eq:rNN} respectively, we interpret the diamond points as (assuming $\sum_{j{=}1}^lq_{j}^+{>}\sum_{j{=}l{+}1}^Lq_{j}^-$, i.e. $\sum_{j{=}1}^La_j{>}\sum_{j{=}1}^Lb_j$ without loss of generality)
\begin{align}
\bar{r}{+}\hat{r}{+}\tilde{r}{=}&L,\\
\bar{r}{+}\hat{r}/2{=}&\sum_{j{=}1}^L\min(a_j{,}b_j){+}\sum_{j{=}l{+}1}^Lq_{j}^-\\
{=}&\sum_{j{=}1}^Lb_j{=}\min(\sum_{j{=}1}^La_j{,}\sum_{j{=}1}^Lb_j),\label{eq:lambdaP}
\end{align}
showing that the corner points are lying on the boundary of inequalities \eqref{eq:theo1}, \eqref{eq:single1} and \eqref{eq:single2}.
\begin{figure}[t]
\renewcommand{\captionfont}{\small}
\captionstyle{center}
\centering \subfigure[$(a_1{,}b_1){=}(0.7{,}0.7)$, $(a_2{,}b_2){=}(0.6{,}0.6)$, $(a_3{,}b_3){=}(0.4{,}0.4)$ and $(a_4{,}b_4){=}(0.3{,}0.3)$]{
                \centering
                \includegraphics[width=0.45\textwidth,height=6.5cm]{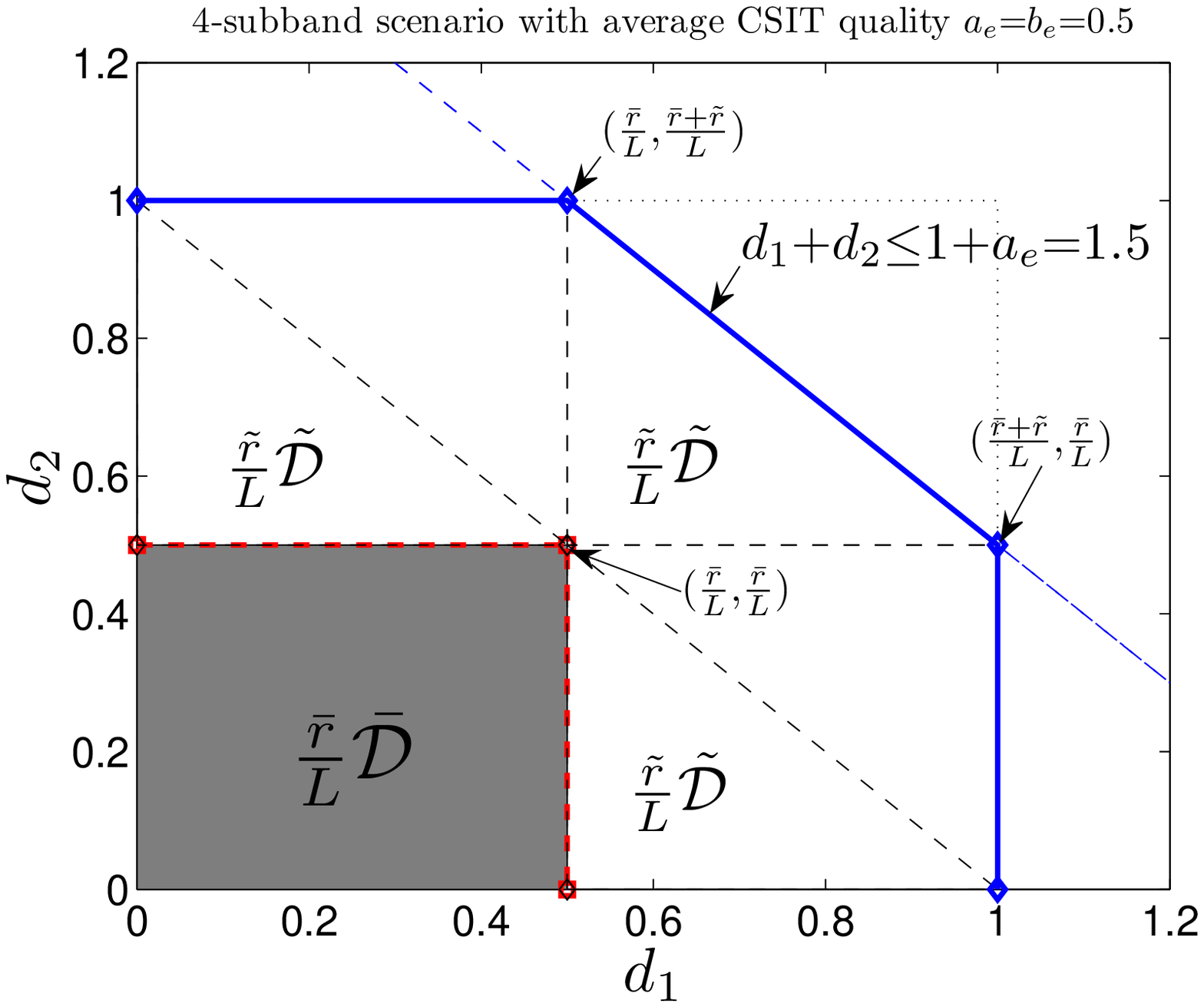}
                \label{fig:D_form2}
        }
\subfigure[$(a_1{,}b_1){=}(1{,}0)$, $(a_2{,}b_2){=}(1{,}0)$, $(a_3{,}b_3){=}(0{,}1)$ and $(a_4{,}b_4){=}(0{,}1)$]{
                \centering
                \includegraphics[width=0.45\textwidth,height=6.5cm]{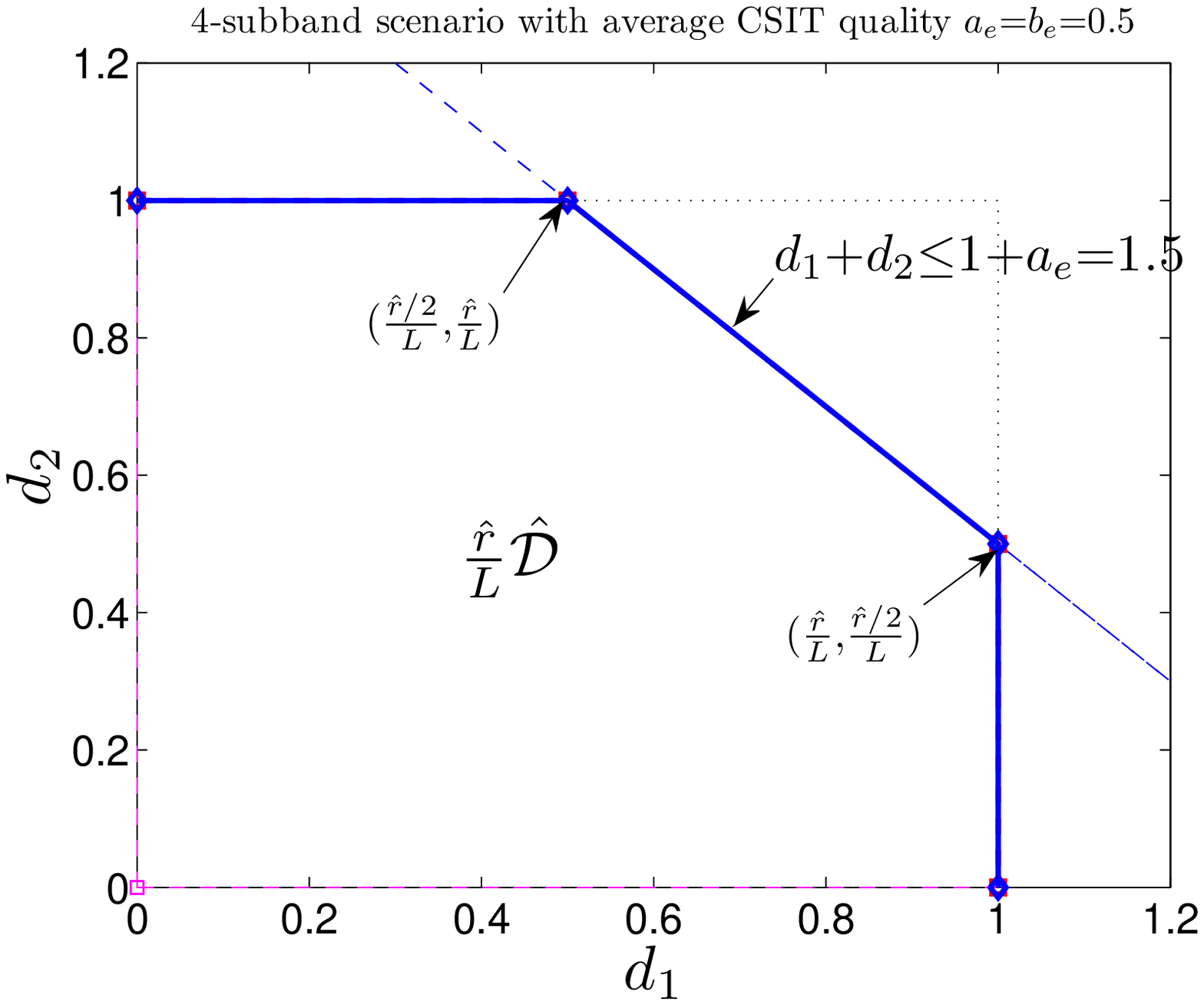}
                \label{fig:D_form3}
        }
\caption{The composition of the optimal \emph{DoF} of a 4-subband scenario, with $(a_e{,}b_e){=}(0.5{,}0.5)$.}\label{fig:ab}
\end{figure}

\newtheorem{myremark}{Remark}
\begin{myremark} \label{rmk:duality}
\textbf{[Equivalence with Remark 1 in \cite{Tandon12}]}

According to \eqref{eq:lambdaP}, the optimal \emph{DoF} region can be rewritten as
\begin{equation}
\mathcal{D}:\quad d_1{+}d_2{\leq}1{+}\frac{1}{L}\min(\sum_{j{=}1}^La_j{,}\sum_{j{=}1}^Lb_j){=}1{+}\frac{\bar{r}{+}\hat{r}/2}{L}.\label{eq:optD2}
\end{equation}
As we have respectively interpreted the weight $\bar{r}$, $\hat{r}$ and $\tilde{r}$ in \eqref{eq:wsum} as the fraction of the subchannels with state $PP$, $NP/PN$ and $NN$, $\frac{\bar{r}{+}\hat{r}/2}{L}$ in \eqref{eq:optD2} in fact stands for the fraction of channel use where the CSIT of a single user is perfect. Revisiting \emph{Remark 1} in \cite{Tandon12}, the sum \emph{DoF} is bounded by
\begin{equation}
d_1{+}d_2{\leq}1{+}\lambda_P{+}\lambda_D,\label{eq:tandon}
\end{equation}
where $\lambda_P$ (resp. $\lambda_D$) refers to the fraction of time where the CSIT of a single user is perfect (resp. delayed). When $\lambda_D{=}0$, \eqref{eq:tandon} becomes a function of $\lambda_P$. Through the weighted-sum interpretation, \eqref{eq:optD2} bridges Theorem \ref{theo1} in this contribution and \emph{Remark 1} in \cite{Tandon12} and find the equivalence between $a_e$ (assuming $a_e{=}b_e$) and $\lambda_P$. Moreover, the upper-bound of the sum \emph{DoF} in \eqref{eq:theo1} generalizes \eqref{eq:tandon} when $\lambda_D{=}0$, because the CSIT states in \cite{Tandon12}, namely $PN$, $NP$, $NN$ and $PP$, are particular cases of the system model investigated in this paper.
\end{myremark}
\begin{myremark} \label{rmk:formation}
\textbf{[The composition of the DoF region changes with CSIT profile]}

Figure \ref{fig:D_form2} and \ref{fig:D_form3} illustrate the formation of the optimal \emph{DoF} region of a 4-subband scenario, with identical average CSIT quality as in Figure \ref{fig:D_form1} (i.e. $a_e{=}b_e{=}0.5$, $L{=}4$), but different profile of the CSIT qualities.

Specifically, in Figure \ref{fig:D_form2}, the transmitter has the knowledge of each user's CSI with the same quality in each subband. The channels are decomposed into subchannels with $PP$ and $NN$ states. The optimal \emph{DoF} region is therefore a function of only $\bar{\mathcal{D}}$ and $\tilde{\mathcal{D}}$. On the other hand, Figure \ref{fig:D_form3} presents an alternating CSIT scenario, whose optimal \emph{DoF} region is composed of only $\hat{\mathcal{D}}$.

Nonetheless, both of these two CSIT settings result in the same \emph{DoF} region as that in Figure \ref{fig:D_form1}. Specifically, in Figure \ref{fig:D_form2}, $\bar{r}{=}\sum_{j{=}1}^4a_j{=}2$ and $\tilde{r}{=}L{-}\bar{r}{=}2$, thus the corner points are $(\frac{1}{2}{,}1)$ and $(1{,}\frac{1}{2})$. While in Figure \ref{fig:D_form3}, $\hat{r}{=}L$ leads to corner points $(\frac{1}{2}{,}1)$ and $(1{,}\frac{1}{2})$. Hence, it is worth noting that if the average CSIT quality per user is fixed, the distribution of the CSIT qualities among the subbands only impacts the composition of the \emph{DoF} region, but does not change the shape.
\end{myremark}
\begin{myremark} \label{rmk:insight_achv}
\textbf{[Relationship between the composition and optimal scheme]}

Since ZFBF, $S_3^{3/2}$ and FDMA are respectively the optimal schemes for the subchannels with state $PP$, $PN/NP$ and $NN$ (as mentioned in Section \ref{dof_subchannel}), the composition of the \emph{DoF} region gives some insights into the optimal transmission scheme.

In Figure \ref{fig:D_form2}, the four subbands are decomposed into subchannels with $PP$ and $NN$ state. The optimality of the \emph{DoF} region is achieved via a scheme integrating ZFBF and FDMA as studied in \cite{pimrc2013}. A similar phenomenon can be observed in Figure \ref{fig:D_form3}. The four subbands merely consists of the subchannels with $PN/NP$ state, whose optimal \emph{DoF} region $\hat{\mathcal{D}}$ composes $\mathcal{D}$ alone. Simply reusing $S_3^{3/2}$ scheme twice (i.e. in subband 1,2 and subband 3, 4), the optimal \emph{DoF} region is achieved.

Moreover, for a scenario with $L{=}2$, $a_1{=}b_2{=}\beta$ and $a_2{=}b_1{=}\alpha$, the subbands are decomposed into $PP$, $PN/NP$ and $NN$ subchannels with $\bar{r}{=}2\alpha$, $\hat{r}{=}2(\beta{-}\alpha)$ and $\tilde{r}{=}2(1{-}\beta)$ channel use respectively (using \eqref{eq:rPP}, \eqref{eq:rPNNP} and \eqref{eq:rNN}). The optimal \emph{DoF} region is achieved via a scheme integrating ZFBF, $S_3^{3/2}$ scheme and FDMA, which is proposed in \cite{Elia13}. Intuitively, the composition of the optimal \emph{DoF} region provides insights into the optimal transmission strategy.
\end{myremark} 

\section{Achievability of $\mathcal{P}_L$ Problem}\label{achp}
In this section, we will discuss the achievability of the optimal \emph{DoF} region for the $\mathcal{P}_L$ problem. We start with evaluating the schemes proposed in \cite{icc13freq}, \cite{Elia13} and \cite{pimrc2013} which investigated a 2-subband scenario with unmatched CSIT (namely $a_1{=}b_2{=}\beta$ and $a_2{=}b_1{=}\alpha$) and matched CSIT (namely, $a_1{=}b_1{=}\beta$ and $a_2{=}b_2{=}\alpha$). By identifying the key ingredients inside the schemes, the optimal scheme for the $\mathcal{P}_L$ problem is found.

\subsection{$\mathcal{P}_1$ Problem}\label{P1}
In \cite{pimrc2013}, a two-subband scenario with matched CSIT (namely $a_1{=}b_1{=}\beta$ and $a_2{=}b_2{=}\alpha$) is studied. As the CSIT quality of each user in each subband is equal to each other ($a_j{=}b_j$), the scenario with matched CSIT can be regarded as two parallel $\mathcal{P}_1$ problems. Reusing the transmission scheme introduced in \cite{pimrc2013}, the optimal \emph{DoF} region can be achieved. For a $\mathcal{P}_1$ problem, the optimal scheme transmits the signal in each subband by superposing a common message I with ZFBF-precoded private messages and writes as
\begin{equation}
\mathbf{x}_1{=}[\underbrace{c_1}_{P{-}P^{a_1}}{,}0]^T{+}\underbrace{\hat{\mathbf{g}}_1^\bot{u_1}}_{P^{a_1}/2}
{+}\underbrace{\hat{\mathbf{h}}_1^\bot{v_1}}_{P^{a_1}/2},\label{eq:P1}
\end{equation}
where $c_1$ is the common message broadcast to both users and $u_1$ and $v_1$ are symbols intended for user 1 and user 2 respectively.

The received signal at each user is expressed as
\begin{align}
y_1{=}\underbrace{h_{11}^*c_1}_{P}{+}\underbrace{\mathbf{h}_1^H\hat{\mathbf{g}}_1^\bot{u_1}}_{P^{a_1}/2}
{+}\underbrace{\mathbf{h}_1^H\hat{\mathbf{h}}_1^\bot{v_1}}_{P^0}{+}\underbrace{\epsilon_{11}}_{P^0},\quad & \quad z_1{=}\underbrace{g_{11}^*c_1}_{P}{+}\underbrace{\mathbf{g}_1^H\hat{\mathbf{g}}_1^\bot{u_1}}_{P^0}
{+}\underbrace{\mathbf{g}_1^H\hat{\mathbf{h}}_1^\bot{v_1}}_{P^{a_1}/2}{+}\underbrace{\epsilon_{12}}_{P^0},
\end{align}
where the private symbols $u_1$ and $v_1$ are drowned by the noise respectively at user 2 and user 1 due to partial ZFBF. Both users decode the common message I first with rate $(1{-}a_1){\log}P$ by treating the private message as noise. Afterwards using Successive Interference Cancelation (SIC), each user can decode their private message with rate $a_1{\log}P$ only subject to noise, after removing the common message. The \emph{DoF} pairs $(1{,}a_1)$ and $(a_1{,}1)$ are achieved if we consider the common message is intended for user 1 and user 2 respectively.

\subsection{$\mathcal{P}_2$ Problem}\label{P2}

As mentioned in Section \ref{P_model}, a $\mathcal{P}_2$ problem considers two basic CSIT quality patterns: 1) $a_1{=}b_1$ and $a_2{=}b_2$; 2) $a_1{\neq}b_1$ and $a_2{\neq}b_2$. The first case is termed as the 2-subband scenario with matched CSIT, which consists of two parallel $\mathcal{P}_1$ problems. The achievability has been discussed in Section \ref{P1}. For the second CSIT quality pattern, the $\mathcal{P}_2$ problem can be considered as the scenario with unmatched CSIT, whose achievable $\emph{DoF}$ region has been investigated in \cite{icc13freq} and \cite{Elia13}. As a reminder, we will identify the shortness of the scheme in \cite{icc13freq} and the benefit of the optimal scheme in \cite{Elia13} through discussion and analysis.

\subsubsection{Optimal Scheme}
\begin{table}[t]
\captionstyle{center} \centering
\renewcommand{\captionfont}{\small}
\begin{tabular}{ccc|ccc}
subband 1& Power & Rate (${\log}P$) & subband 2 & Power & Rate (${\log}P$)\\
$c_1$ & $P{-}P^{a_1}$ & $1{-}a_1$ & $c_2$ & $P{-}P^{b_2}$ & $1{-}b_2$\\
$u_1$ & $P^{b_1}/2$ & $b_1$ & $u_2$ & $P^{b_2}/2$ & $b_2$\\
$u_0$ & $(P^{a_1}-P^{b_1})/2$ & $a_1{-}b_1$ & $u_0$ & $(P^{b_2}-P^{a_2})/2$ & $b_2{-}a_2$\\
$v_1$ & $P^{a_1}/2$ & $a_1$ & $v_2$ & $P^{a_2}/2$ & $a_2$
\end{tabular}
\caption{Power and rate allocation in the optimal scheme for $\mathcal{P}_2$ problem.}\label{tab:power_rate}
\end{table}
The optimal transmission blocks in \emph{subband $1$} and $2$ are expressed as
\begin{align}
\mathbf{x}_1&= \left[c_1,0\right]^T{+}\hat{\mathbf{g}}_1^\bot u_1{+}[u_0{,}0]^T{+}\hat{\mathbf{h}}_1^\bot{v_1},\label{eq:opt_x1}\\
\mathbf{x}_2&= \left[c_2,0\right]^T{+}\hat{\mathbf{h}}_2^\bot v_2{+}[u_0{,}0]^T{+}\hat{\mathbf{g}}_2^\bot{u_2}.\label{eq:opt_x2}
\end{align}
Common message II, $u_0$, and common messages I, $c_1$ and $c_2$ should be decoded by both users (but could be intended for user 1 and user 2 respectively or exclusively for user 1 or user 2). Note that we do not precode common messages I and II in this paper as it does not impact the \emph{DoF}. $u_1$ and $u_2$ are symbols intended for user 1, while $v_1$ and $v_2$ are symbols intended for user 2. The rate and power allocation are shown in Table \ref{tab:power_rate}, resulting in the following received signals at each user
\begin{align}
y_1{=}&\underbrace{h_{11}^*c_1}_P{+}\underbrace{\mathbf{h}_1^H\hat{\mathbf{g}}_1^\bot u_1}_{P^{b_1}}{+}\underbrace{h_{11}^*u_0}_{P^{a_1}}{+}
\underbrace{\mathbf{h}_1^H\hat{\mathbf{h}}_1^\bot{v_1}}_{P^0}{+}\epsilon_{11},\label{eq:opty1}\\
z_1{=}&\underbrace{g_{11}^*c_1}_P{+}\underbrace{\mathbf{g}_1^H\hat{\mathbf{g}}_1^\bot
u_1}_{P^0}{+}\underbrace{g_{11}^*u_0}_{P^{a_1}}{+}
\underbrace{\mathbf{g}_1^H\hat{\mathbf{h}}_1^\bot{v_1}}_{P^{b_1}}{+}\epsilon_{12},\label{eq:optz1}\\
y_2{=}&\underbrace{h_{21}^*c_2}_P{+}\underbrace{\mathbf{h}_2^H\hat{\mathbf{g}}_2^\bot{u_2}}_{P^{b_2}}{+}\underbrace{h_{21}^*u_0}_{P^{b_2}}{+}
\underbrace{\mathbf{h}_2^H\hat{\mathbf{h}}_2^\bot v_2}_{P^0}{+}\epsilon_{21},\label{eq:opty2}\\
z_2{=}&\underbrace{g_{21}^*c_2}_P{+}\underbrace{\mathbf{g}_2^H\hat{\mathbf{g}}_2^\bot{u_2}}_{P^0}{+}\underbrace{g_{21}^*u_0}_{P^{b_2}}
{+}\underbrace{\mathbf{g}_2^H\hat{\mathbf{h}}_2^\bot v_2}_{P^{a_2}}{+}\epsilon_{22}.\label{eq:optz2}
\end{align}

In \eqref{eq:opty1}, \eqref{eq:optz1} and \eqref{eq:opty2}, \eqref{eq:optz2}, $c_1$ and $c_2$ are respectively decoded first by treating all the other terms as noise. Afterwards, user 1 decodes $u_0$ and $u_1$ from $y_1$ using SIC. With the knowledge of $u_0$, $u_2$ can be recovered from $y_2$. Similarly, user 2 decodes $u_0$ and $v_2$ from $z_2$ via SIC. $v_1$ can be decoded from $z_1$ by eliminating $u_0$.

To keep the same notation as in \cite{Elia13,icc13freq,pimrc2013}, we replace $a_1{,}b_2$ with $\beta$ and $a_2{,}b_1$ with $\alpha$ and $\beta{\geq}\alpha$. The \emph{DoF} pair $(1{,}\frac{a_1{+}a_2}{2}){=}(1{,}\frac{\alpha{+}\beta}{2})$ and $(\frac{a_1{+}a_2}{2}{,}1){=}(\frac{\alpha{+}\beta}{2}{,}1)$ are achieved if we consider the common messages are intended for user 1 and user 2 respectively, consistent with the optimal \emph{DoF} region. Note that when $\beta{=}\alpha$, the $\mathcal{P}_2$ problem will degrade to two parallel $\mathcal{P}_1$ problems and no common message II, $u_0$, is generated.

\subsubsection{Shortness of the Scheme Proposed in \cite{icc13freq}}\label{shortness}
In order to identify the shortness of the suboptimal scheme, we keep the same notation as in \cite{Elia13,icc13freq,pimrc2013}, namely $a_1{=}b_2{=}\beta$ and $a_2{=}b_1{=}\alpha$ and $\beta{\geq}\alpha$. In the suboptimal scheme, the transmit signals in \emph{subband $1$} and $2$ are respectively expressed as
\begin{align}
\mathbf{x}_1 &{=}\left[c_1{,}0\right]^T\!\!\!{+}\left[\mu_1{,}0\right]^T\!\!\!{+}
[\hat{\mathbf{h}}_1^\bot{,}\hat{\mathbf{h}}_1][v_{11}{,}v_{12}]^T{+}\hat{\mathbf{g}}_1^\bot{u_1},\label{eq:sA2}\\
\mathbf{x}_2 &{=}\left[c_2{,}0\right]^T\!\!\!{+}\left[\mu_2,{0}\right]^T\!\!\!{+}
[\hat{\mathbf{g}}_2^\bot{,}\hat{\mathbf{g}}_2][u_{21}{,}u_{22}]^T{+}\hat{\mathbf{h}}_2^\bot{v_2},\label{eq:sB2}
\end{align}
where the private symbols $u_1$, $v_{11}$, $u_{21}$ and $v_2$ are precoded and transmitted with the power and rate similar to $u_1$, $v_1$, $u_2$ and $v_2$ in \eqref{eq:opt_x1} and \eqref{eq:opt_x2} respectively.

Besides, $v_{12}$ and $u_{22}$, generated with rate $(\beta{-}\alpha){\log}P$, are respectively overheard by user 1 in subband 1 and by user 2 in subband 2, thus leading to the requirement of transmitting $\mu{=}v_{12}{+}u_{22}$ to enable the decoding of other private symbols. $\mu$ is further split into $\mu_1$ and $\mu_2$ and multicast via an extra $\beta{-}\alpha$ channel use. However, no extra channel use is required in the optimal scheme, because $u_0$ is sent twice (i.e. subband 1 and 2) so that each user can decode it alternatively.

To sum up, the scheme in \cite{icc13freq} employs $2\beta{+}\beta{-}\alpha$ channel use to transmit six private symbols (i.e. $v_{11}$, $v_{12}$, $u_1$, $u_{21}$, $u_{22}$, $v_2$), while the optimal scheme sends five symbols (i.e. $u_1$, $v_2$, $u_0$, $v_1$, $u_2$) in $2\beta$ channel use. Besides, the common messages, $c_1$ and $c_2$, are sent using ${\max}(2{-}3\beta{+}\alpha{,}0)$ and $2{-}2\beta$ channel use in the sub-optimal and optimal scheme respectively. Their sum \emph{DoF} are respectively expressed as
\begin{align}
d_\Sigma^{sub}{=}&\frac{2\beta{+}2\alpha{+}2(\beta{-}\alpha){+}{\max}(2{-}3\beta{+}\alpha{,}0)}
{3\beta{-}\alpha{+}{\max}(2{-}3\beta{+}\alpha{,}0)}\label{eq:dof_subopt1}\\
{=}&\frac{4\alpha{+}4(\beta{-}\alpha){+}{\max}(2{-}3\beta{+}\alpha{,}0)}{2\alpha{+}3(\beta{-}\alpha){+}{\max}(2{-}3\beta{+}\alpha{,}0)},\label{eq:dof_subopt}\\
d_\Sigma^{opt}{=}&\frac{2\beta{+}2\alpha{+}(\beta{-}\alpha){+}2{-}2\beta}{2\beta{+}2{-}2\beta}\label{eq:dof_opt1}\\
{=}&\frac{4\alpha{+}3(\beta{-}\alpha){+}2{-}2\beta}{2\alpha{+}2(\beta{-}\alpha){+}2{-}2\beta}.\label{eq:dof_opt}
\end{align}
\begin{myremark}\label{rmk:compare}
\textbf{[Shortness of the suboptimal scheme]}

The sum \emph{DoF} performance is further derived as \eqref{eq:dof_subopt} and \eqref{eq:dof_opt}, which provide an explicit interpretation of the sub-optimality of \cite{icc13freq}. More precisely, with the weighted-sum interpretation, the denominator and nominator in each equation are written as the sum of three parts. Specifically, in \eqref{eq:dof_opt}, $2\alpha$ channel use is employed by ZFBF and $4\alpha{\log}P$ sum rate (namely the rate of $u_1$, $v_2$ and part of the rate of $u_2$ and $v_1$) is achieved; $S_3^{3/2}$ scheme performs on $2(\beta{-}\alpha)$ channel use and achieves $3(\beta{-}\alpha){\log}P$ sum rate (namely the rate of $u_0$ and part of the rate of $u_2$ and $v_1$); The common messages, $c_1$ and $c_2$, are sent via FDMA with $2{-}2\beta$ channel use.

However, as in \eqref{eq:dof_subopt}, the scheme in \cite{icc13freq} combines ZFBF, MAT and FDMA. $2\alpha$ channel use is employed by ZFBF and $4\alpha{\log}P$ sum rate (namely the rate of $u_1$, $v_2$ and part of the rate of $u_{21}$ and $v_{11}$) is achieved; MAT scheme performs on $3(\beta{-}\alpha)$ channel use and achieves $4(\beta{-}\alpha){\log}P$ sum rate (namely the rate of $v_{12}$ and $u_{22}$ and part of the rate of $u_{21}$ and $v_{11}$). Compared to $S_3^{3/2}$, MAT scheme employs an extra $\beta{-}\alpha$ channel use, but only results in $\beta{-}\alpha$ rate improvement. At the same time, the fraction of FDMA transmission (namely, $c_j$) is shrunk to $2{-}3\beta{+}\alpha$. A \emph{DoF} loss is incurred when $2{-}3\beta{+}\alpha{<}0$, because the transmission of $\mu_1$ and $\mu_2$ (requiring an extra $\beta{-}\alpha$ channel use) cannot be completed using the remaining channel use (after generating the private symbols, which is $2{-}2\beta$) in subband 1 and 2.
\end{myremark}

\begin{figure}[t]
\renewcommand{\captionfont}{\small}
\centering
\includegraphics[height=5cm,width=8cm]{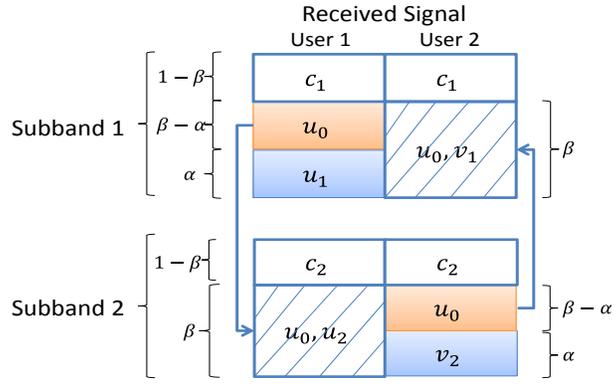}
\caption{The illustration of the received signal and decoding procedure of the optimal scheme for the $\mathcal{P}_2$ problem with $a_1{=}b_2{=}\beta$, $a_2{=}b_1{=}\alpha$ and $\beta{\geq}\alpha$, where the values beside the the bracket stand for the (pre-log factor of the) rate of the corresponding symbols. User 1 (resp. user 2) observes $u_0$ with higher power than $u_1$ (resp. $v_2$) in subband 1 (resp. 2) and receives $u_0$ with the same power level as $u_2$ (resp. $v_1$) in subband 2 (resp. 1). The common message $u_0$ can be decoded by both users but in different subbands. Then each user employs it to eliminate the interference and decode the private symbols.}\label{fig:s3}
\end{figure}

\begin{myremark}\label{rmk:key_u0}
\textbf{[Key point in the optimal scheme]}

Figure \ref{fig:s3} provides an illustrative description of the received signals and decoding procedure of the optimal scheme. The key point to boost the \emph{DoF} lies in making both users decode $u_0$ without the employment of any extra channel use. To this end, the transmitter broadcasts $u_0$ twice, i.e. subband 1 and 2. In subband 1, user 1 is said to be more capable to decode $u_0$ because it receives $u_0$ with a higher power than the private symbol $u_1$. Similarly, user 2 is more capable in subband 2. In this way, both users can decode $u_0$ alternatively when they are more capable and no extra channel use is required.

As we will see in the following two subsections, this insight can be generalized to solve $\mathcal{P}_L{,}L{\geq}3$ problem by generating multiple streams of $u_0$ and sending each of them twice. One is in subband $j_1{\in}[1{:}l]$ and the other is in subband $j_2{\in}[l{+}1{,}L]$, where user 1 and user 2 are respectively more capable to decode $u_0$.
\end{myremark}

\subsection{$\mathcal{P}_3$ Problem}\label{P3}
\begin{table}[t]
\captionstyle{center} \centering
\renewcommand{\captionfont}{\small}
\begin{tabular}{ccc|ccc|ccc}
& Power & Rate (${\log}P$) & & Power & Rate (${\log}P$) & & Power & Rate (${\log}P$)\\
$c_1$ & $P{-}P^{a_1}$ & $1{-}a_1$ & $c_2$ & $P{-}P^{a_2}$ & $1{-}a_2$ & $c_3$ & $P{-}P^{b_3}$ & $1{-}b_3$\\
$u_1$ & $P^{b_1}/2$ & $b_1$ & $u_2$ & $P^{b_2}/2$ & $b_2$ & $u_3$ & $P^{b_3}/2$ & $b_3$\\
$v_1$ & $P^{a_1}/2$ & $a_1$ & $v_2$ & $P^{a_2}/2$ & $a_2$ & $v_3$ & $P^{a_3}/2$ & $a_3$\\
$u_0(1)$ & $P^{a_1}/2{-}P^{b_1}/2$ & $a_1{-}b_1$ & $u_0(2)$ & $P^{a_2}/2{-}P^{b_2}/2$ & $a_2{-}b_2$ & $u_0(1)$ & $P^{b_3{-}q_2^+}/2{-}P^{a_3}/2$ & $q_1^+{=}a_1{-}b_1$\\
& & & & & & $u_0(2)$ & $P^{b_3}/2{-}P^{b_3{-}q_2^+}/2$ & $q_2^+{=}b_2{-}a_2$
\end{tabular}
\caption{Power and rate allocation in the optimal scheme for 3-subband case.}\label{tab:power_rate3}
\end{table}
In this part, we investigate a $\mathcal{P}_3$ problem ($\sum_{j{=}1}^3a_j{=}\sum_{j{=}1}^3b_j$) with $a_1{\geq}b_1$, $a_2{\geq}b_2$, $a_3{\leq}b_3$ without the loss of generality. Inspired by Remark \ref{rmk:key_u0}, we construct the optimal transmission block as follows
\begin{align}
\mathbf{x}_1=&\left[c_1,0\right]^T{+}\hat{\mathbf{g}}_1^\bot u_1{+}[u_{0}(1){,}0]^T{+}\hat{\mathbf{h}}_1^\bot{v_1},\label{eq:sb3x1}\\
\mathbf{x}_2=&\left[c_2,0\right]^T{+}\hat{\mathbf{g}}_2^\bot u_2{+}[u_{0}(2){,}0]^T{+}\hat{\mathbf{h}}_2^\bot{v_2},\label{eq:sb3x2}\\
\mathbf{x}_3=&\left[c_3,0\right]^T{+}\hat{\mathbf{g}}_3^\bot u_3{+}[u_{0}(2){+}u_{0}(1){,}0]^T{+}\hat{\mathbf{h}}_3^\bot{v_3},\label{eq:sb3x3}
\end{align}
where $u_{0}(1)$, $u_{0}(2)$, $c_1$, $c_2$ and $c_3$ are common messages, $u_j$ and $v_j$ ($j{=}1{,}2{,}3$) are private symbols respectively intended for user 1 and 2. The power and rate allocation are given in Table \ref{tab:power_rate3}. As presented, $u_{0}(1)$ and $u_{0}(2)$ are respectively sent in subband 1 and 2 when user 1 is more capable to decode them since $a_1{>}b_1$ and $a_2{>}b_2$. In subband 3, $u_0(1)$ and $u_0(2)$ are transmitted again via superposition coding and user 2 has the capability to decode both of them as $q_3^-{=}q_1^+{+}q_2^+$. The received signals at user 1 and user 2 are expressed as
\begin{align}
y_1{=}&\underbrace{h_{11}^*c_1}_P{+}\underbrace{\mathbf{h}_1^H\hat{\mathbf{g}}_1^\bot u_1}_{P^{b_1}}{+}\underbrace{h_{11}^*u_{0}(1)}_{P^{a_1}}{+}
\underbrace{\mathbf{h}_1^H\hat{\mathbf{h}}_1^\bot{v_1}}_{P^0}{+}\epsilon_{11},\label{eq:sb3y1}\\
z_1{=}&\underbrace{g_{11}^*c_1}_P{+}\underbrace{\mathbf{g}_1^H\hat{\mathbf{g}}_1^\bot
u_1}_{P^0}{+}\underbrace{g_{11}^*u_{0}(1)}_{P^{a_1}}{+}
\underbrace{\mathbf{g}_1^H\hat{\mathbf{h}}_1^\bot{v_1}}_{P^{a_1}}{+}\epsilon_{12},\label{eq:sb3z1}\\
y_2{=}&\underbrace{h_{21}^*c_2}_P{+}\underbrace{\mathbf{h}_2^H\hat{\mathbf{g}}_2^\bot u_2}_{P^{b_2}}{+}\underbrace{h_{21}^*u_{0}(2)}_{P^{a_2}}{+}
\underbrace{\mathbf{h}_2^H\hat{\mathbf{h}}_2^\bot{v_2}}_{P^0}{+}\epsilon_{21},\label{eq:sb3y2}\\
z_2{=}&\underbrace{g_{21}^*c_2}_P{+}\underbrace{\mathbf{g}_2^H\hat{\mathbf{g}}_2^\bot u_2}_{P^0}{+}\underbrace{g_{21}^*u_{0}(2)}_{P^{a_2}}
{+}\underbrace{\mathbf{g}_2^H\hat{\mathbf{h}}_2^\bot{v_2}}_{P^{a_2}}{+}\epsilon_{22},\label{eq:sb3z2}\\
y_3{=}&\underbrace{h_{31}^*c_3}_P{+}\underbrace{\mathbf{h}_3^H\hat{\mathbf{g}}_3^\bot u_3}_{P^{b_3}}{+}h_{31}^*(\underbrace{u_{0}(2)}_{P^{b_3}}{+}\underbrace{u_{0}(1)}_{P^{b_3{-}q_2^+}}){+}
\underbrace{\mathbf{h}_3^H\hat{\mathbf{h}}_3^\bot{v_3}}_{P^0}{+}\epsilon_{31},\label{eq:sb3y3}\\
z_3{=}&\underbrace{g_{31}^*c_3}_P{+}\underbrace{\mathbf{g}_3^H\hat{\mathbf{g}}_3^\bot u_3}_{P^0}{+}g_{31}^*(\underbrace{u_{0}(2)}_{P^{b_3}}{+}\underbrace{u_{0}(1)}_{P^{b_3{-}q_2^+}}){+}
\underbrace{\mathbf{g}_3^H\hat{\mathbf{h}}_3^\bot{v_3}}_{P^{a_3}}{+}\epsilon_{32},\label{eq:sb3z3}
\end{align}
\begin{figure}[t]
\renewcommand{\captionfont}{\small}
\centering
\includegraphics[height=7cm,width=8cm]{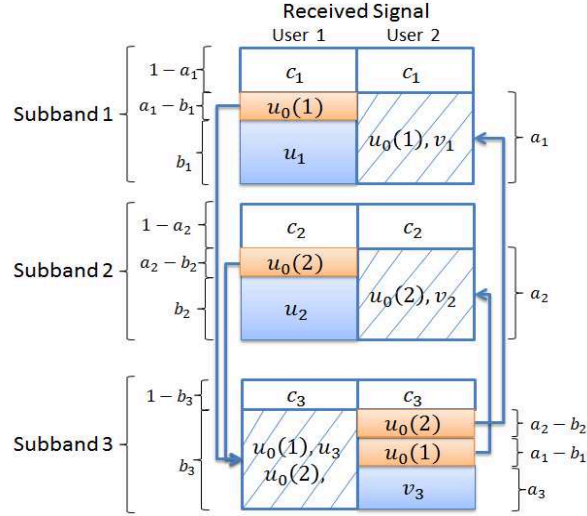}
\caption{The illustration of the received signal and decoding procedure of the optimal scheme for the $\mathcal{P}_3$ problem, where the values beside the the bracket stand for the (pre-log factor of the) rate of the corresponding symbols. $u_0(1)$ and $u_0(2)$ are transmitted. User 1 decodes them in subband 1 and 2 respectively. User 2 recovers them using SIC in subband 3.}\label{fig:P3}
\end{figure}

\emph{Decoding:} At both users, the common messages $c_1$, $c_2$ and $c_3$ are respectively decoded from the observation in subband 1, 2 and 3. After that, at user 1, $u_{0}(1)$ and $u_{0}(2)$ are respectively decoded from $y_1$ and $y_2$ by treating $u_1$ and $u_2$ as noise, since $a_1$ and $a_2$ are respectively greater than $b_1$ and $b_2$. With the knowledge of $u_{0}(1)$ and $u_{0}(2)$, the private symbols $u_1$, $u_2$ and $u_3$ are obtained from $y_1$, $y_2$ and $y_3$ respectively using SIC.

At user 2, treating $u_{0}(1)$ and $v_3$ as noise, $u_{0}(2)$ is decoded from $z_3$ with the SNR as
\begin{equation}
SNR_{u_{0}(2)}{\approx}\frac{P^{b_3}}{P^{b_3{-}q_2^+}}{=}P^{q_2^+}{=}P^{a_2{-}b_2},\text{when } P\to\infty.
\end{equation}
Removing $u_{0}(2)$ and treating $v_3$ as noise, $u_{0}(1)$ is decoded with the SNR as
\begin{equation}
SNR_{u_{0}(1)}{\approx}\frac{P^{b_3{-}q_2^+}}{P^{a_3}}{=}P^{b_3{-}q_2^+{-}a_3}{=}P^{q_1^+}{=}P^{a_1{-}b_1},\text{when } P\to\infty.
\end{equation}
With the knowledge of $u_{0}(1)$ and $u_{0}(2)$, the private symbols $v_1$, $v_2$ and $v_3$ can be decoded from $z_1$, $z_2$ and $z_3$ respectively using SIC.

The private symbols $u_1{,}u_2{,}u_3$ intended for user 1 achieve the sum rate $(b_1{+}b_2{+}b_3){\log}P$, so do the private symbols $v_1{,}v_2{,}v_3$ for user 2 since $\sum_{j{=}1}^3a_j{=}\sum_{j{=}1}^3b_j$. Considering the common messages intended for user 1 and user 2 respectively, we have the \emph{DoF} pair $(1{,}\frac{1}{3}\sum_{j{=}1}^3b_j)$ and $(\frac{1}{3}\sum_{j{=}1}^3a_j{,}1)$.

It is worth noting that for the scenario with $a_1{>}b_1$, $a_2{=}b_2$ and $a_3{<}b_3$, the problem turns to a combination of one $\mathcal{P}_1$ problem (i.e. subband 2) and one $\mathcal{P}_2$ problem (i.e. subband 1 and 3). Specifically, the transmitted signal in subband 2 become exactly the same as that in \eqref{eq:P1} because $u_0(2)$ is not generated. The transmitted signals in subband 1 and 3 follow the form as in \eqref{eq:opt_x1} and \eqref{eq:opt_x2} and $u_0(1)$ is the only common message II to be sent. Moreover, for the case $a_1{=}b_1$, $a_2{=}b_2$ and $a_3{=}b_3$, the transmitted signals in \eqref{eq:sb3x1} to \eqref{eq:sb3x3} degrade to \eqref{eq:P1} and no common messages II is generated.

\begin{myremark}\label{rmk:P3_u0_decomp}
\textbf{[Insights behind the solution to $\mathcal{P}_3$]}

The construction of the transmission block (shown from \eqref{eq:sb3x1} to \eqref{eq:sb3x3}) relates to the channel decomposition discussed in Section \ref{wsum}. $q_1^+$ and $q_2^+$ represent the fraction of channel use of the subchannels with $PN$ state and $q_3^-$ stands for that of $NP$ state. In order to find the alternating $PN/NP$ scenario, we recall from the further decomposition of subchannels with $PN$ and $NP$ states as in Section \ref{dof_subchannel} and in Figure \ref{fig:further_decomp}. Subchannel $\hat{3}$ is further decomposed into $\hat{3}(1)$ and $\hat{3}(2)$, with channel use $q_1^+$ and $q_2^+$ respectively. Consequently, subchannels $\hat{1}$ and $\hat{3}(1)$, subchannels $\hat{2}$ and $\hat{3}(2)$ are paired. $u_0(1)$ and $u_0(2)$ are respectively the transmissions performed on those pairs of subchannels. Figure \ref{fig:P3} illustrates the philosophy of decoding. As shown, $u_0(1)$ and $u_0(2)$ act as two separated and independent layers of the common messages. User 1 (in subband 1 and 2) and user 2 (in subband 3) are alternatively capable to decode them. Hence, the strategy in solving a $\mathcal{P}_3$ problem is an extension of that solving a $\mathcal{P}_2$ problem.
\end{myremark}

\subsection{$\mathcal{P}_L$ Problem}\label{PL}

We build the optimal transmission block for the $\mathcal{P}_L$ problem following the discussion on the $\mathcal{P}_3$ problem. Briefly, the private symbols in each subband is transmitted using ZFBF precoding. The rate and power allocated to them are functions of the quality of the CSIT of their unintended user. Afterwards, every common message II, namely $u_0(\cdot)$ symbol, is generated based on the insight discussed in Remark \ref{rmk:P3_u0_decomp} and transmitted through one antenna. Finally, common message I in each subband (i.e. $c_j$) is transmitted through a single antenna via the remaining channel use. The procedure of generating the transmission signal is sketched below
\begin{enumerate}
  \item In each subband, generate the private symbols $u_j$ and $v_j$ respectively with the power $P^{b_j}$ and $P^{a_j}$ and rate $b_j{\log}P$ and $a_j{\log}P$, ${\forall}j{\in}[1{,}L]$.
  \item $i{\leftarrow}1$; If $\{q^+\}$ or $\{q^-\}$ has all zero elements, goto Step 7), otherwise, goto Step 3).
  \item Arbitrarily pair subbands $j_1{\in}[1{:}l]$ and $j_2{\in}[l{+}1{:}L]$, such that $q_{j_1}^+{\neq}0$ and $q_{j_2}^-{\neq}0$.
  \item Generate common message II, $u_{0}(i)$, with rate $\min(q_{j_1}^+{,}q_{j_2}^-){\log}P$ and transmit it in subband $j_1$ and $j_2$.
  \item If $q_{j_1}^+{<}q_{j_2}^-$, update $q_{j_2}^-{\leftarrow}q_{j_2}^-{-}q_{j_1}^+$ and $q_{j_1}^+{\leftarrow}0$; Else if $q_{j_1}^+{>}q_{j_2}^-$, update $q_{j_1}^+{\leftarrow}q_{j_1}^+{-}q_{j_2}^-$ and $q_{j_2}^-{\leftarrow}0$; Else if $q_{j_1}^+{=}q_{j_2}^-$, update $q_{j_1}^+{\leftarrow}0$ and $q_{j_2}^-{\leftarrow}0$.
  \item $i{\leftarrow}i{+}1$; If $\{q^+\}$ or $\{q^-\}$ has all zero elements, goto Step 7), otherwise, goto Step 3).
  \item For the subbands with $a_j{<}1$ and $b_j{<}1$, generate common message I, $c_j$, with rate $(1{-}\max(a_j{,}b_j)){\log}P$ and power $P{-}P^{\max(a_j{,}b_j)}$.
\end{enumerate}
\begin{figure}[t]
\renewcommand{\captionfont}{\small}
\centering
\includegraphics[height=8cm,width=14cm]{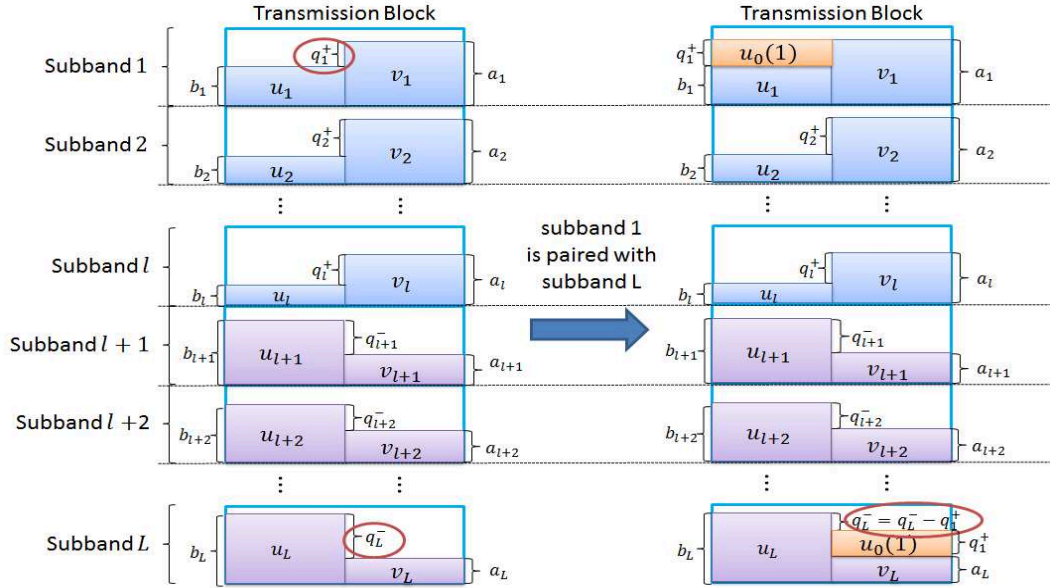}
\caption{The illustration of the generation of the optimal transmission block for $\mathcal{P}_L$ problem. The values beside the brackets represent the (pre-log factor of the) rate allocated to the corresponding symbols.}\label{fig:PL}
\end{figure}

Figure \ref{fig:PL} illustrates the generation of the transmission block for the $\mathcal{P}_L$ problem. As shown, the private symbols are generated following Step 1). After that, subband 1 and subband $L$ are paired as in Step 3), in each of which $u_0(1)$ is generated and transmitted with rate $q_1^+{\log}P$ following Step 4). $q_1^+$ becomes zero and $q_3^-$ turns to $q_3^-{-}q_1^+$ according to Step 5). Keep generating $u_0(i)$ messages following Step 3) to 5) until either the set $\{q^+\}$ or $\{q^-\}$ has all zero elements.

Consequently, the transmit signal in general consists of a common message I ($c_j$), ZFBF-precoded private symbols ($u_j$ and $v_j$) and superposition-coded multiple common messages II ($u_0(\cdot)$) symbols. It writes as
\begin{equation}
\mathbf{x}_j{=}[c_j{,}0]^T{+}\hat{\mathbf{g}}_j^\bot{u_j}{+}\hat{\mathbf{h}}_j^\bot{v_j}{+}
[\sum_{i{=}1}^{K_j{=}|\mathcal{K}_j|}u_{0}(\mathcal{K}_j(i)){,}0]^T,\label{eq:sbLxj}
\end{equation}
where $\mathcal{K}_j$, with the cardinality $K_j$, is the set of the $u_0(\cdot)$ symbols to be sent in subband $j$. The power and rate allocation for the symbols transmitted in subband $j{\in}[1{,}l]$ are presented in Table \ref{tab:power_rateL}, where $\tau_j(i)$ represents the rate of $u_0(\mathcal{K}_j(i))$. Also, we have $P^{a_j{-}\sum_{i{=}1}^{K_j}\tau_j(i)}{=}P^{b_j}$, namely $\sum_{i{=}1}^{K_j}\tau_j(i){=}a_j{-}b_j$, such that all the $u_0(\cdot)$ symbols in the set $\mathcal{K}_j$ can be recovered.
\begin{table}[t]
\captionstyle{center} \centering
\renewcommand{\captionfont}{\small}
\begin{tabular}{ccc}
& Power & Rate (${\log}P$)\\
$c_j$ & $P{-}P^{a_j}$ & $1{-}a_j$ \\
$u_j$ & $P^{b_j}/2$ & $b_j$\\
$v_j$ & $P^{a_j}/2$ & $a_j$\\
$u_0(\mathcal{K}_j(1))$ & $(P^{a_j}{-}P^{a_j{-}\tau_j(1)})/2$ & $\tau_j(1)$\\
$u_0(\mathcal{K}_j(2))$ & $(P^{a_j{-}\tau_j(1)}{-}P^{a_j{-}\tau_j(1){-}\tau_j(2)})/2$ & $\tau_j(2)$\\
$\vdots$ & $\vdots$ & $\vdots$\\
$u_0(\mathcal{K}_j(K_j))$ & $(P^{a_j{-}\sum_{i{=}1}^{K_j{-}1}\tau_j(i)}{-}P^{a_j{-}\sum_{i{=}1}^{K_j}\tau_j(i)})/2$ & $\tau_j(K_j)$
\end{tabular}
\caption{Power and rate allocation in the optimal scheme for $\mathcal{P}_L$ problem in the subband with $j{\in}[1{,}l]$.}\label{tab:power_rateL}
\end{table}

The signal received at each receiver in subband $j{\leq}l$ is expressed as
\begin{align}
y_j{=}&\underbrace{h_{j1}^*c_j}_{P}{+}\underbrace{\mathbf{h}_j^H\hat{\mathbf{g}}_j^\bot{u_j}}_{P^{b_j}}
{+}\underbrace{\mathbf{h}_j^H\hat{\mathbf{h}}_j^\bot{v_j}}_{P^0}
{+}h_{j1}^*(\underbrace{u_0(\mathcal{K}_j(1))}_{P^{a_j}}{+}\underbrace{u_0(\mathcal{K}_j(2))}_{P^{a_j{-}\tau_j(1)}}{+}\cdots
{+}\underbrace{u_0(\mathcal{K}_j(K_j))}_{P^{a_j{-}\sum_{k{=}1}^{K_j{-}1}\tau_j(k)}}){+}\epsilon_{j1},\label{eq:sbLyl}\\
z_j{=}&\underbrace{g_{j1}^*c_j}_{P}{+}\underbrace{\mathbf{g}_j^H\hat{\mathbf{g}}_j^\bot{u_j}}_{P^0}
{+}\underbrace{\mathbf{g}_j^H\hat{\mathbf{h}}_j^\bot{v_j}}_{P^{a_j}}
{+}g_{j1}^*(\underbrace{u_0(\mathcal{K}_j(1))}_{P^{a_j}}{+}\underbrace{u_0(\mathcal{K}_j(2))}_{P^{a_j{-}\tau_j(1)}}{+}\cdots
{+}\underbrace{u_0(\mathcal{K}_j(K_j))}_{P^{a_j{-}\sum_{k{=}1}^{K_j{-}1}\tau_j(k)}}){+}\epsilon_{j2}.\label{eq:sbLzl}
\end{align}

\emph{Decoding:} At both users, $c_j$ can be decoded first by treating all the other terms as noise. After removing $c_j$, user 1 sees $u_0(\mathcal{K}_j(i)){,}i{=}1{,}2{,}{\cdots}{,}K_j$ with different power levels and decodes them using SIC. Specifically, $u_0(\mathcal{K}_j(i)){,}i{<}K_j$ is decoded with the SNR as
\begin{equation}
SNR_{u_0(\mathcal{K}_j(i))}{\approx}\frac{P^{a_j{-}\sum_{i^\prime{=}1}^{i{-}1}\tau_j(i^\prime)}}{P^{a_j{-}\sum_{i^\prime{=}1}^i\tau_j(i^\prime)}}
{=}P^{\tau_j(i)},\text{when } P{\to}{\infty}.
\end{equation}
By treating $u_j$ as noise, $u_0(\mathcal{K}_j(K_j))$ is recovered with the SNR as
\begin{equation}
SNR_{u_0(\mathcal{K}_j(K_j))}{\approx}\frac{P^{a_j{-}\sum_{i^\prime{=}1}^{K_j{-}1}\tau_j(i^\prime)}}{P^{b_j}}
{=}P^{\tau_j(i)},\text{when } P{\to}{\infty},
\end{equation}
since $\sum_{i{=}1}^{K_j}\tau_j(i){=}a_j{-}b_j$. After removing $u_0(\mathcal{K}_j(i)){,}i{=}1{,}2{,}{\cdots}{,}K_j$ from $y_j$, user 1 recovers $u_j$ subject to noise.

Performing the same decoding procedure for subbands $1$ to $l$ (with $a_j{\geq}b_j$), user 1 can recover every $u_0(\cdot)$ symbol. However in the subbands $l{+}1$ to $L$ (with $a_j{\leq}b_j$), user 1 sees a mixture of the $u_0(\cdot)$ message and $u_j$. Since every $u_0(\cdot)$ symbol is recovered from $y_1$ to $y_l$, the private symbols intended for user 1 in subbands $l{+}1$ to $L$ are recovered with the knowledge of all the $u_0(\cdot)$ symbols. User 2 can decode its messages similarly.

The sum rate achieved by the private symbols, $u_{1{:}L}$, intended for user 1, is $\sum_{j{=}1}^Lb_j{\log}P$. The private symbols, $v_{1{:}L}$, intended for user 2, achieve the sum rate $\sum_{j{=}1}^La_j{\log}P$. Besides, the common messages I, $c_{1{:}L}$, achieve the sum rate $(L{-}\sum_{j{=}1}^L{\max}(a_j{,}b_j)){\log}P$. Combined with the sum rate of common messages II ($u_0(\cdot)$), namely $\sum_{j{=}1}^lq_j^+{\log}P{=}\frac{1}{2}\sum_{j{=}1}^L|a_j{-}b_j|$, the sum rate of all the symbols is $(L{+}\sum_{j{=}1}^La_j){\log}P$. If all the common messages (i.e. $u_0(\cdot)$ symbols and $c_{1{:}L}$) are intended for user 1, the \emph{DoF} pair $(1{,}\frac{1}{L}\sum_{j{=}1}^La_j)$ is achieved, thus solving the $\mathcal{P}_L$ problem.

\begin{myremark}\label{rmk:weight_rate}
\textbf{[Rate allocation and weights calculation]}

Relating the rate of the symbols presented in Table \ref{tab:power_rateL} to the weights in \eqref{eq:wsum}, we find that the weighted-sum interpretation of the \emph{DoF} region reveals not only the integration of FDMA, $S_3^{3/2}$ scheme and ZFBF, but also the rate allocation. To be specific, the common messages $c_j$ are sent via FDMA. The sum rate of all common messages $c_j$ is consistent with $\tilde{r}$, the fraction of channel use of the subchannels with $NN$ state. Moreover, private symbols $u_{1{:}l}$ and $v_{l{+}1{:}L}$ are transmitted via ZFBF, their sum rate is equal to the weight of $PP$ state, $\bar{r}$. The sum rate of all the $u_0(\cdot)$ messages reflects the fraction of channel use of subchannels with $PN$ or $NP$ state.
\end{myremark}

\section{Achievability of $\mathcal{Q}_L$ Problem}\label{achq}
In this section, we will focus on the category of $\mathcal{Q}_{L}$ problem and work out the optimal scheme that achieves the \emph{DoF} region specified in Theorem \ref{theo1}. Without loss of generality, we only investigate the $\mathcal{Q}_L^+$ problem as $\mathcal{Q}_L^-$ can be solved by simply switching the role of the two users.

\subsection{$\mathcal{Q}_1^+$ Problem}\label{Q1}
As a reminder, in the $\mathcal{Q}_1^+$ problem, we have $a_1{>}b_1$. The optimal transmission strategy is straightforward and identical to that in \eqref{eq:P1} by substituting $a_1$ with $a_1^\prime{=}b_1$. Specifically, the power and rate allocated to the private symbol $v_1$ is $P^{a_1^\prime}$ and $a_1^\prime{\log}P$. Since $a_1^\prime{<}a_1$, $v_1$ is drowned by the noise in the observation of user 1. Performing the same decoding procedure as in the $\mathcal{P}_1$ problem, the \emph{DoF} pair $(1{,}a_1^\prime){=}(1{,}b_1)$ and $(b_1{,}1)$ can be achieved.
\begin{myremark}\label{rmk:Q1}
\textbf{[$\mathcal{P}_1$ and $\mathcal{Q}_1$]}

We can conclude that the $\mathcal{Q}_1^+$ problem with $a_1{>}b_1$ is equivalent to the $\mathcal{P}_1$ problem with $a_1^\prime{=}b_1$ in terms of \emph{DoF} region and the optimal transmission strategy. In other words, for a $\mathcal{Q}_1^+$ problem, the CSIT of user 1 is over-accurate compared to the CSIT qualities of user 2 and does not enhance the \emph{DoF} region. We can also observe this using Corollary \ref{corollary_wsum}. The \emph{DoF} region of a $\mathcal{Q}_1^+$ problem and a $\mathcal{P}_1$ problem have exactly the same weighted-sum interpretation. To be specific, after the subband in a $\mathcal{Q}_1^+$ problem is decomposed, the $\hat{r}^\prime{=}a_1{-}b_1$ channel use of the subchannel with singular $PN$ state is merged with the subchannel with $NN$ state. In this way, the channel use of $NN$ state is $1{-}b_1$, identical to that in a $\mathcal{P}_1$ problem with $a_1^\prime{=}b_1$.
\end{myremark}

\subsection{$\mathcal{Q}_2^+$ Problem}\label{Q2}
\begin{figure}[t]
\renewcommand{\captionfont}{\small}
\centering
\includegraphics[height=5cm,width=13cm]{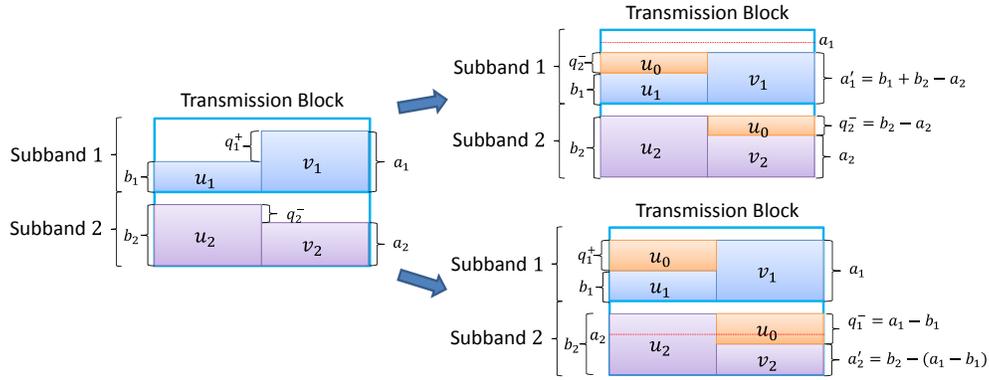}
\caption{The illustration of the generation of the optimal transmission block for the $\mathcal{Q}_2$ problem, where the values beside the bracket stand for the (pre-log factor of the) rate of the corresponding symbols. Two rate and power allocation policies are presented.}\label{fig:Q2}
\end{figure}
Similarly to the scenario considered in a $\mathcal{P}_2$ problem, there exists two basic scenario in a $\mathcal{Q}_2^+$ problem, namely 1) $a_1{\geq}b_1$ and $a_2{\geq}b_2$; 2) $a_1{\geq}b_1$ and $a_2{<}b_2$. The first case can be regarded as two $\mathcal{Q}_1^+$ problems or a $\mathcal{Q}_1^+$ problem with a $\mathcal{P}_1$ problem. The achievability has been studied in Section \ref{Q1} and Section \ref{P1}. For the second CSIT quality pattern, the optimal scheme is designed by reusing the philosophy of the transmission strategy discussed in Section \ref{P2}. The challenge lies in the power and rate allocation for $u_0$ and private symbols.

For concreteness, we initially allocate the rate of the private symbols $u_j$ as $b_j{\log}P$ and $v_j$ as $a_j{\log}P$. Reusing the transmission in \eqref{eq:opt_x1} and \eqref{eq:opt_x2}, user 1 could decode $u_0$ with rate $q_1^+{\log}P$ but user 2 could do $q_2^-{\log}P$. Hence, to make $u_0$ decodable by both users, there exist two options:
\begin{enumerate}
\item Determine the rate of $u_0$ as $q_2^-{\log}P{=}(b_2{-}a_2){\log}P$ and decrease the rate of $v_1$ to $a_1^\prime{\log}P{=}(b_1{+}q_2^-){\log}P$ (from $a_1{\log}P$);
\item Generate $u_0$ with the rate $q_1^+{\log}P{=}(a_1{-}b_1){\log}P$ and reduce the rate of $v_2$ to $a_2^\prime{\log}P{=}(b_2{-}q_1^+){\log}P$ (from $a_2{\log}P$).
\end{enumerate}
Figure \ref{fig:Q2} gives an illustration of these two constructions of the transmit signal. The transmitted signals write as in \eqref{eq:opt_x1} and \eqref{eq:opt_x2}, but the power and rate allocation are changed and shown in Table \ref{tab:power_rateq2}.
\begin{table}[t]
\captionstyle{center} \centering
\renewcommand{\captionfont}{\small}
\begin{tabular}{c|ccc|ccc}
Option 1 & \emph{subband $1$} & Power & Rate (${\log}P$) & \emph{subband $2$} & Power & Rate (${\log}P$)\\
&$c_1$ & $P{-}P^{q_2^-{+}b_1}$ & $1{-}q_2^-{-}b_1$ & $c_2$ & $P{-}P^{b_2}$ & $1{-}b_2$\\
&$u_1$ & $P^{b_1}/2$ & $b_1$ & $u_2$ & $P^{b_2}/2$ & $b_2$\\
&$v_1$ & $P^{q_2^-{+}b_1}/2$ & ${q_2^-{+}b_1}$ & $v_2$ & $P^{a_2}/2$ & $a_2$\\
&$u_0$ & $P^{q_2^-{+}b_1}/2{-}P^{b_1}/2$ & $q_2^-$ & $u_0$ & $P^{b_2}/2{-}P^{a_2}/2$ & $q_2^-$\\
\hline Option 2 & \emph{subband $1$} & Power & Rate (${\log}P$) & \emph{subband $2$} & Power & Rate (${\log}P$)\\
&$c_1$ & $P{-}P^{a_1}$ & $1{-}a_1$ & $c_2$ & $P{-}P^{q_1^+{+}a_2}$ & $1{-}q_1^+{-}a_2$\\
&$u_1$ & $P^{b_1}/2$ & $b_1$ & $u_2$ & $P^{b_2}/2$ & $b_2$\\
&$v_1$ & $P^{a_1}/2$ & $a_1$ & $v_2$ & $P^{b_2{-}q_1^+}/2$ & $b_2{-}q_1^+$\\
&$u_0$ & $P^{a_1}/2{-}P^{b_1}/2$ & $q_1^+$ & $u_0$ & $P^{b_2}/2{-}P^{b_2{-}q_1^+}/2$ & $q_1^+$
\end{tabular}
\caption{Power and rate allocation in the optimal scheme for the $\mathcal{Q}_2$, where $q_2^-{=}b_2{-}a_2$ and $q_1^+{=}a_1{-}b_1$.}\label{tab:power_rateq2}
\end{table}

Employing the first power and rate allocation policy, the signal received at each user can be written as
\begin{align}
y_1{=}&\underbrace{h_{11}^*c_1}_P{+}\underbrace{\mathbf{h}_1^H\hat{\mathbf{g}}_1^\bot {u_1}}_{P^{b_1}}{+}\underbrace{h_{11}^*u_0}_{P^{q_2^-{+}b_1}}{+}
\underbrace{\mathbf{h}_1^H\hat{\mathbf{h}}_1^\bot{v_1}}_{P^{q_2^-{+}b_1}P^{{-}a_1}{<}P^0}{+}\epsilon_{11},\label{eq:q2y1_1}\\
z_1{=}&\underbrace{g_{11}^*c_1}_P{+}\underbrace{\mathbf{g}_1^H\hat{\mathbf{g}}_1^\bot
{u_1}}_{P^0}{+}\underbrace{g_{11}^*u_0}_{P^{q_2^-{+}b_1}}{+}
\underbrace{\mathbf{g}_1^H\hat{\mathbf{h}}_1^\bot{v_1}}_{P^{q_2^-{+}b_1}}{+}\epsilon_{12},\label{eq:q2z1_1}\\
y_2{=}&\underbrace{h_{21}^*c_2}_P{+}\underbrace{\mathbf{h}_2^H\hat{\mathbf{g}}_2^\bot{u_2}}_{P^{b_2}}{+}\underbrace{h_{21}^*u_0}_{P^{b_2}}{+}
\underbrace{\mathbf{h}_2^H\hat{\mathbf{h}}_2^\bot v_2}_{P^0}{+}\epsilon_{21},\label{eq:q2y2_1}\\
z_2{=}&\underbrace{g_{21}^*c_2}_P{+}\underbrace{\mathbf{g}_2^H\hat{\mathbf{g}}_2^\bot{u_2}}_{P^0}{+}\underbrace{g_{21}^*u_0}_{P^{b_2}}
{+}\underbrace{\mathbf{g}_2^H\hat{\mathbf{h}}_2^\bot v_2}_{P^{a_2}}{+}\epsilon_{22},\label{eq:q2z2_1}
\end{align}
The decoding procedure is the same as in $\mathcal{P}_2$ problem discussed in Section \ref{P2}. Both users' private symbols ($u_1{,}u_2$ and $v_1{,}v_2$) achieve the sum rate $(q_2^-{+}b_1{+}a_2){\log}P{=}(b_1{+}b_2){\log}P$. Besides, the common messages $c_1$, $c_2$ and $u_0$, achieving the sum rate $R_{c_1}{+}R_{c_2}{+}R_{u_0}{=}(2{-}b_1{-}b_2){\log}P$, can be considered as exclusively intended for user 1 or user 2. As a consequence, the \emph{DoF} pair $(1{,}(b_1{+}b_2)/2){=}(1{,}\min(a_e{,}b_e))$ and $((b_1{+}b_2)/2{,}1){=}(\min(a_e{,}b_e){,}1)$ are achieved.

With the second power and rate allocation policy, the received signals write as
\begin{align}
y_1{=}&\underbrace{h_{11}^*c_1}_P{+}\underbrace{\mathbf{h}_1^H\hat{\mathbf{g}}_1^\bot {u_1}}_{P^{b_1}}{+}\underbrace{h_{11}^*u_0}_{P^{a_1}}{+}
\underbrace{\mathbf{h}_1^H\hat{\mathbf{h}}_1^\bot{v_1}}_{P^0}{+}\epsilon_{11},\label{eq:q2y1_2}\\
z_1{=}&\underbrace{g_{11}^*c_1}_P{+}\underbrace{\mathbf{g}_1^H\hat{\mathbf{g}}_1^\bot
{u_1}}_{P^0}{+}\underbrace{g_{11}^*u_0}_{P^{a_1}}{+}
\underbrace{\mathbf{g}_1^H\hat{\mathbf{h}}_1^\bot{v_1}}_{P^{a_1}}{+}\epsilon_{12},\label{eq:q2z1_2}\\
y_2{=}&\underbrace{h_{21}^*c_2}_P{+}\underbrace{\mathbf{h}_2^H\hat{\mathbf{g}}_2^\bot{u_2}}_{P^{b_2}}{+}\underbrace{h_{21}^*u_0}_{P^{b_2}}{+}
\underbrace{\mathbf{h}_2^H\hat{\mathbf{h}}_2^\bot v_2}_{P^{b_2{-}q_1^+}P^{-a_2}{<}P^0}{+}\epsilon_{21},\label{eq:q2y2_2}\\
z_2{=}&\underbrace{g_{21}^*c_2}_P{+}\underbrace{\mathbf{g}_2^H\hat{\mathbf{g}}_2^\bot{u_2}}_{P^0}{+}\underbrace{g_{21}^*u_0}_{P^{b_2}}
{+}\underbrace{\mathbf{g}_2^H\hat{\mathbf{h}}_2^\bot v_2}_{P^{b_2{-}q_1^+}}{+}\epsilon_{22}.\label{eq:q2z2_2}
\end{align}
Performing the same decoding procedure, the same rate pair is achieved.
\begin{myremark}\label{rmk:Q2}
\textbf{[$\mathcal{P}_2$ and $\mathcal{Q}_2$, extension of Remark \ref{rmk:Q1}]}

A noteworthy observation from the above scheme is that a $\mathcal{Q}_2$ problem can be considered as equivalent to a $\mathcal{P}_2$ problem. Specifically, the first power and rate allocation strategy is identical to the optimal scheme for a $\mathcal{P}_2$ problem with $a_1^\prime{=}b_1{+}b_2{-}a_2$ while fixing $b_1{,}b_2$ and $a_2$. The second allocation policy coincides with the optimal scheme for a $\mathcal{P}_2$ problem with $a_2^\prime{=}b_2{-}(a_1{-}b_1)$ while fixing $b_1{,}b_2$ and $a_1$. This equivalence is due to the fact that the difference between the CSIT quality of user 1 and user 2 (i.e. $a_1{+}a_2{-}b_1{-}b_2$), interpreted as the singular $\hat{r}^\prime$ $PN$ state according to the discussion in Section \ref{wsum}, does not benefit the \emph{DoF} region more than a $NN$ state based on Theorem \ref{theo1}.
\end{myremark}

\subsection{$\mathcal{Q}_L^+$ Problem}\label{QL}
\begin{figure}[t]
\renewcommand{\captionfont}{\small}
\centering
\includegraphics[height=3cm,width=15cm]{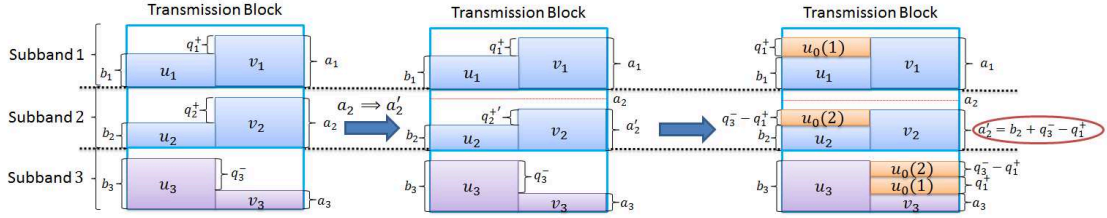}
\caption{The illustration of the generation of the transmission block of the optimal scheme for the $\mathcal{Q}_3$ problem, where the values beside the the bracket stand for the (pre-log factor of the) rate of the corresponding symbols. The $\mathcal{Q}_3$ problem is transformed to a $\mathcal{P}_3$ problem and solved.}\label{fig:Q3}
\end{figure}
Following Remark \ref{rmk:Q1} and \ref{rmk:Q2}, to design the optimal scheme for the $\mathcal{Q}_L^+$ problem, we establish a $\mathcal{P}_L$ problem with $\{a_1^\prime{,}a_2^\prime{,}\cdots{,}a_L^\prime\}$ and $\{b_1{,}b_2{,}\cdots{,}b_L\}$, such that
\begin{align}
\sum_{j{=}1}^La_j^\prime{=}&\sum_{j{=}1}^Lb_j{,}\\
a_j^\prime{\leq}&a_j{,}{\forall}j{\in}[1{,}L].
\end{align}

After that, the optimal transmission block for the $\mathcal{Q}_L^+$ problem is constructed following the footsteps presented in Section \ref{PL}. Here we do not rewrite the transmitted and received signals as they are similar to that in Section \ref{PL}. Instead, we only explain how the corner points, $(1{,}b_e)$ and $(b_e{,}1)$, are achieved.

According to Step 1) in the procedure given in Section \ref{PL}, the power allocated to $v_j$ (the private symbol intended for user 2 in subband $j$) scales as $P^{a_j^\prime}$. This does not introduce interference to the received signal at user 1 since $\mathbf{h}_j^H\hat{\mathbf{h}}_j^\bot{\sim}P^{-a_j}$ and $a_j^\prime{\leq}a_j$. Therefore, the sum rate of $u_{1{:}L}$ and $v_{1{:}L}$ respectively become $\sum_{j{=}1}^Lb_j{\log}P$ and $\sum_{j{=}1}^La_j^\prime{\log}P$. The common messages II, $u_0(\cdot)$, are generated following Step 3) to 6). The sum rate of $u_0(\cdot)$ is $(\sum_{a_j^\prime{>}b_j}a_j^\prime{-}b_j){\log}P$ because all the common messages are transmitted twice, i.e. once in the subbands with $a_j^\prime{>}b_j$ and once in the subbands with $b_j{>}a_j^\prime$. According to Step 7), common messages I, $c_{1{:}L}$ achieve the rate of $(L{-}\sum_{j{=}1}^L{\max}(a_j^{\prime}{,}b_j)){\log}P$. Considering all the common messages I and II are intended for user 1, we have $d_1$ computed as
\begin{align}
d_1{=}&\frac{1}{L}(\sum_{j{=}1}^Lb_j{+}\sum_{a_j^\prime{>}b_j}a_j^\prime{-}b_j{+}L{-}\sum_{j{=}1}^L{\max}(a_j^{\prime}{,}b_j))\\
{=}&\frac{1}{L}(\sum_{j{=}1}^Lb_j{+}\sum_{a_j^\prime{\geq}b_j}a_j^\prime{-}b_j{+}L
{-}\sum_{a_j^\prime{\geq}b_j}a_j^\prime{-}\sum_{a_j^\prime{<}b_j}b_j)\\
{=}&\frac{1}{L}(\sum_{a_j^\prime{\geq}b_j}b_j{+}\sum_{a_j^\prime{<}b_j}b_j{+}\sum_{a_j^\prime{\geq}b_j}a_j^\prime{-}
\sum_{a_j^\prime{\geq}b_j}b_j{+}L{-}\sum_{a_j^\prime{\geq}b_j}a_j^\prime{-}\sum_{a_j^\prime{<}b_j}b_j)\label{eq:exp2}\\
{=}&1.
\end{align}
\eqref{eq:exp2} follows the fact that $\sum_{j{=}1}^Lb_j{=}\sum_{a_j^\prime{\geq}b_j}b_j{+}\sum_{a_j^\prime{<}b_j}b_j$. Hence, the \emph{DoF} pair $(d_1{,}d_2){=}(1{,}\frac{1}{L}\sum_{j{=}1}^La_j^\prime){=}(1{,}b_e)$ is achieved. Similarly, assuming the common messages I and II are intended for user 2, the \emph{DoF} pair $(b_e{,}1)$ is achieved.

For concreteness, we consider a $\mathcal{Q}_3^+$ problem with $a_1{>}b_1$, $a_2{>}b_2$ and $a_3{<}b_3$, namely $q_1^+{+}q_2^+{>}q_3^-$. As shown in Figure \ref{fig:Q3}, the construction of the transmission block is obtained by establishing a $\mathcal{P}_3$ problem with $a_1^\prime{=}a_1$, $a_2^\prime{=}b_1{+}b_2{+}b_3{-}a_1{-}a_3$ and $a_3^\prime{=}a_3$. Besides, we assume $a_2^\prime{>}0$ and $a_2^\prime{>}b_2$. The transmitted signals write as
\begin{align}
\mathbf{x}_1=&\underbrace{\left[c_1,0\right]^T}_{P{-}P^{a_1}}{+}\underbrace{\hat{\mathbf{g}}_1^\bot{u_1}}_{P^{b_1}/2}{+}
\underbrace{\left[u_{0}(1),0\right]^T}_{(P^{a_1}{-}P^{b_1})/2}{+}\underbrace{\hat{\mathbf{h}}_1^\bot{v_1}}_{P^{a_1}/2},\label{eq:sb3x1Q3}\\
\mathbf{x}_2=&\underbrace{\left[c_2,0\right]^T}_{P{-}P^{a_2^\prime}}{+}\underbrace{\hat{\mathbf{g}}_2^\bot{u_2}}_{P^{b_2}/2}{+}
\underbrace{\left[u_{0}(2),0\right]^T}_{(P^{a_2^\prime}{-}P^{b_2})/2}{+}
\underbrace{\hat{\mathbf{h}}_2^\bot{v_2}}_{P^{a_2^\prime}/2},\label{eq:sb3x2Q3}\\
\mathbf{x}_3=&\underbrace{\left[c_3,0\right]^T}_{P{-}P^{b_3}}{+}\underbrace{\hat{\mathbf{g}}_3^\bot{u_3}}_{P^{b_3}/2}{+}
\underbrace{\left[u_0(2){,}0\right]^T}_{(P^{b_3}{-}P^{b_3{-}q_3^-{+}q_1^+})/2}{+}
\underbrace{\left[u_0(1){,}0\right]^T}_{(P^{b_3{-}q_3^-{+}q_1^+}{-}P^{a_3})/2}{+}
\underbrace{\hat{\mathbf{h}}_3^\bot{v_3}}_{P^{a_3}/2}.\label{eq:sb3x3Q3}
\end{align}

The received signal can be derived similarly as that in \eqref{eq:sb3y1} to \eqref{eq:sb3z3}. The decoding procedure follows as that in $\mathcal{P}_3$ problem (see Figure \ref{fig:P3}). Generally, both user recover $c_1$, $c_2$ and $c_3$ from their observations by treating all the other terms as noise. Then user 1 can decode the $u_0(1)$ and $u_0(2)$ from $y_1$ and $y_2$ (the subbands with $a_j{>}b_j$) respectively while user 2 recovers them from $z_3$ (the subbands with $a_j{<}b_j$). With the knowledge of $u_0(1)$ and $u_0(2)$, all the private messages are decoded.

The private symbols $u_1{,}u_2{,}u_3$ achieve the sum rate $(b_1{+}b_2{+}b_3){\log}P$ while the private symbols $v_1{,}v_2{,}v_3$ achieve the sum-rate $(a_1{+}a_2^\prime{+}a_3){\log}P$. Considering that $c_1$, $c_2$, $c_3$, $u_0(1)$ and $u_0(2)$ are intended for user 1 and user 2, the \emph{DoF} pair $(1{,}\frac{1}{3}(b_1{+}b_2{+}b_3))$ and $(\frac{1}{3}(a_1{+}a_2^\prime{+}a_3){,}1)$ are respectively obtained.

\begin{myremark}\label{rmk:tradeoff}
\textbf{[Transmission Strategies vs. Feedback Quality Distribution]}

For an $L$-subband scenario with given $a_e$ and $b_e$, the optimal transmission strategy proposed in Section \ref{PL} and \ref{QL} provide a general solution to achieve the optimal \emph{DoF} region. However, the form of the transmitted signal in each subband varies depending on the CSIT quality pattern.

For instance, with CSIT quality pattern $a_j{\geq}b_j{,}{\forall}j{\in}[1{:}L]$, the $L$-subband scenario comprises $L$ times of $\mathcal{P}_1$ or $\mathcal{Q}_1$ problems. The transmitted signal in each subband is independent to each other and no common messages II, namely $u_0(\cdot)$, is generated. Moreover, with per-user average CSIT quality $a_e{=}b_e$, an extreme scenario is defined as $(a_j{,}b_j){=}(1{,}1){,}{\forall}j{\in}[1{,}a_e{\times}L]$ and $(a_j{,}b_j){=}(0{,}0){,}{\forall}j{\in}[a_e{\times}L{+}1{,}L]$, which means that the CSIT states are $PP$ for subbands 1 to $a_eL$ while they are $NN$ for the remaining subbands. Hence, the optimal schemes become ZFBF in subband 1 to $a_eL$ while the optimal schemes in the remaining subbands degrade to FDMA.

For a $\mathcal{P}_L$ problem with the CSIT quality pattern satisfying the condition that ${\forall}\mathcal{J}{\subset}[1{:}L]{,}\sum_{j{\in}\mathcal{J}}a_j{\neq}\sum_{j{\in}\mathcal{J}}b_j$, there are totally $L{-}1$ $u_0(\cdot)$ symbols generated and the transmitted signal in each subband is correlated to each other. Besides, there are multiple stages in the SIC and the decoding of private symbols rely on the $u_0(\cdot)$ symbols.

\end{myremark}


\section{Conclusion} \label{conclusion}

In this contribution, we investigate a general two-user frequency correlated MISO BC, which consists of multiple subbands with varying CSIT qualities. A tight outer-bound to the \emph{DoF} region is found with the help of Nair-Gamal's bound \cite{NairGamal}, Extremal Inequality \cite{Extremal} and Lemma 1 in \cite{Ges12}. Its optimality is shown by a transmission block as an extension of the optimal scheme for a two-subband scenario. Due to the varying CSIT qualities, the two users are alternatively capable of decoding the common messages $u_0(\cdot)$. To achieve the optimal \emph{DoF} performance, the number of the common messages $u_0(\cdot)$ and the rate of each $u_0(\cdot)$ message are determined accordingly. It is worth noting that the optimal \emph{DoF} region is a function of the minimum average CSIT quality between the two users. This result provides confirmative answer to the conjecture made in \cite{Tandon12} that the \emph{DoF} region in the two-user MISO BC with perfect CSIT of only one user is the same as that with no CSIT of either user.

This optimal \emph{DoF} region is interpreted as the weighted-sum of the optimal \emph{DoF} region in the CSIT states $PP$, $PN/NP$ and $NN$. The weight of each CSIT state is calculated according to the CSIT qualities of both users in each subband and indicates the fraction of channel use of each type of CSIT states. For a fixed per-user average CSIT, the distribution of the CSIT qualities of each user across the $L$ subbands only impacts the formation of the optimal \emph{DoF} region, but does not influence the shape of the region. This sheds light on the construction of the optimal transmission scheme.

\section*{Appendix-Derivation of \eqref{eq:appendix}}
To obtain \eqref{eq:appendix}, we introduce
\begin{align}
\Phi_j{=}&\Phi_{j}^{(1)}{+}\Phi_{j}^{(2)},\text{for }j{=}1{,}2{,}{\cdots}{,}\lfloor\frac{n{+}1}{2}\rfloor,\label{eq:phij}\\
\Theta_j{=}&\Theta_{j}^{(1)}{+}\Theta_{j}^{(2)},\text{for }j{=}1{,}2{,}{\cdots}{,}n,\label{eq:thetaj}
\end{align}
where
\begin{align}
\Phi_{j}^{(1)}{=}&I(M_1{,}Z_{n{-}j{+}2}^n;Y_{j}^{n{-}j{+}1}|\mathcal{S}_1^n{,}Y_1^{j{-}1}),\\
\Phi_{j}^{(2)}{=}&I(M_2;Z_{j}^{n{-}j{+}1}|M_1{,}\mathcal{S}_1^n{,}Y_1^{j{-}1}{,}Z_{n{-}j{+}2}^n),\\
\Theta_{j}^{(1)}{=}&I(M_1{,}Y_1^{j{-}1}{,}Z_{j{+}1}^n{,}\hat{\mathcal{S}}_1^n;Y_j|\mathcal{S}_1^n),\\
\Theta_{j}^{(2)}{=}&I(M_2{,}Y_1^{j{-}1}{,}Z_{j{+}1}^n{,}\hat{\mathcal{S}}_1^n;Z_j|M_1{,}\mathcal{S}_1^n{,}Y_1^{j{-}1}{,}Z_{j{+}1}^n).
\end{align}

Consequently, \eqref{eq:appendix} can be rewritten as
\begin{equation}
\Phi_1{\leq}\sum_{j{=}1}^n\Theta_{j}.\label{eq:appendix2}
\end{equation}

In order to show that \eqref{eq:appendix2} holds, we need the following upper-bounds of $\Phi_j{,}{\forall}j{\leq}\lfloor\frac{n{+}1}{2}\rfloor$, namely
\begin{align}
\Phi_{j{<}\lfloor\frac{n{+}1}{2}\rfloor}{\leq}&\Theta_j{+}\Theta_{n{-}j{+}1}{+}\Phi_{j{+}1},\label{eq:recursive1}\\
\Phi_{\lfloor\frac{n{+}1}{2}\rfloor}{\leq}&\left\{\begin{array}{ll}\Theta_{\lfloor\frac{n{+}1}{2}\rfloor}
{+}\Theta_{\lfloor\frac{n{+}1}{2}\rfloor{+}1},&\text{if n is even;}\\ \Theta_{\lfloor\frac{n{+}1}{2}\rfloor},&\text{if n is odd.}\end{array}\right.\label{eq:recursive2}
\end{align}
Using \eqref{eq:recursive1} and \eqref{eq:recursive2}, the summation of $\Phi_j{,}{\forall}j{\leq}\lfloor\frac{n{+}1}{2}\rfloor$ is bounded as
\begin{align}
\sum_{j{=}1}^{\lfloor\frac{n{+}1}{2}\rfloor{-}1}\Phi_j{+}\Phi_{\lfloor\frac{n{+}1}{2}\rfloor}{\leq}&
\sum_{j{=}1}^{\lfloor\frac{n{+}1}{2}\rfloor{-}1}\left\{\Theta_j{+}\Theta_{n{-}j{+}1}{+}\Phi_{j{+}1}\right\}
{+}\left\{\begin{array}{ll}\Theta_{\lfloor\frac{n{+}1}{2}\rfloor}
{+}\Theta_{\lfloor\frac{n{+}1}{2}\rfloor{+}1},&\text{if n is even;}\\ \Theta_{\lfloor\frac{n{+}1}{2}\rfloor},&\text{if n is odd}\end{array}\right.\\
{=}&\sum_{j{=}2}^{\lfloor\frac{n{+}1}{2}\rfloor}\Phi_j
{+}\underbrace{\sum_{j{=}1}^{\lfloor\frac{n{+}1}{2}\rfloor{-}1}\left\{\Theta_j{+}\Theta_{n{-}j{+}1}\right\}
{+}\left\{\begin{array}{ll}\Theta_{\lfloor\frac{n{+}1}{2}\rfloor}
{+}\Theta_{\lfloor\frac{n{+}1}{2}\rfloor{+}1},&\text{if n is even;}\\ \Theta_{\lfloor\frac{n{+}1}{2}\rfloor},&\text{if n is odd}\end{array}\right.}_{\sum_{j{=}1}^n\Theta_j}
\end{align}
Eliminating $\Phi_2{,}\Phi_3{,}{\cdots}\Phi_{\lfloor\frac{n{+}1}{2}\rfloor}$ in both l.h.s and r.h.s, \eqref{eq:appendix2} holds. Next, we aim at showing \eqref{eq:recursive1} and \eqref{eq:recursive2}.

\subsection{When $j{<}\lfloor\frac{n{+}1}{2}\rfloor$}
$\Phi_j^{(1)}$ is derived as follows.
\begin{align}
\Phi_{j}^{(1)}{=}&
I(M_1{,}Z_{n{-}j{+}2}^n;Y_{n{-}j{+}1}|\mathcal{S}_1^n{,}Y_1^{n{-}j}){+}I(M_1{,}Z_{n{-}j{+}2}^n;Y_{j}^{n{-}j}|\mathcal{S}_1^n{,}Y_1^{j{-}1})\label{eq:phi11}\\
{\leq}&I(M_1{,}Z_{n{-}j{+}2}^n{,}Y_1^{n{-}j};Y_{n{-}j{+}1}|\mathcal{S}_1^n){+}I(M_1{,}Z_{n{-}j{+}2}^n;Y_{j}|\mathcal{S}_1^n{,}Y_1^{j{-}1}){+}
I(M_1{,}Z_{n{-}j{+}2}^n;Y_{j{+}1}^{n{-}j}|\mathcal{S}_1^n{,}Y_1^{j})\label{eq:remcondy}\\
{=}&I(M_1{,}Z_{n{-}j{+}2}^n{,}Y_1^{n{-}j};Y_{n{-}j{+}1}|\mathcal{S}_1^n)\nonumber\\
&{+}I(M_1{,}Z_{j{+}1}^n;Y_{j}|\mathcal{S}_1^n{,}Y_1^{j{-}1}){-}I(Z_{j{+}1}^{n{-}j{+}1};Y_{j}|\mathcal{S}_1^n{,}M_1{,}Z_{n{-}j{+}2}^n{,}Y_1^{j{-}1})\nonumber\\
&{+}I(M_1{,}Z_{n{-}j{+}1}^n;Y_{j{+}1}^{n{-}j}|\mathcal{S}_1^n{,}Y_1^{j}){-}
I(Z_{n{-}j{+}1};Y_{j{+}1}^{n{-}j}|\mathcal{S}_1^n{,}M_1{,}Z_{n{-}j{+}2}^n{,}Y_1^{j})\label{eq:remcondy1}\\
{\leq}&\underbrace{I(M_1{,}Z_{n{-}j{+}2}^n{,}Y_1^{n{-}j}{,}\hat{\mathcal{S}}_1^n;Y_{n{-}j{+}1}|\mathcal{S}_1^n)}_{\Theta_{n{-}j{+}1}^{(1)}}{+}
\underbrace{I(M_1{,}Z_{j{+}1}^n{,}Y_1^{j{-}1}{,}\hat{\mathcal{S}}_1^n;Y_{j}|\mathcal{S}_1^n)}_{\Theta_{j}^{(1)}}\nonumber\\
&{+}\underbrace{I(M_1{,}Z_{n{-}j{+}1}^n;Y_{j{+}1}^{n{-}j}|\mathcal{S}_1^n{,}Y_1^{j})}_{\Phi_{j{+}1}^{(1)}}\nonumber\\
&{-}I(Z_{j{+}1}^{n{-}j{+}1};Y_{j}|\mathcal{S}_1^n{,}M_1{,}Z_{n{-}j{+}2}^n{,}Y_1^{j{-}1}){-}
I(Z_{n{-}j{+}1};Y_{j{+}1}^{n{-}j}|\mathcal{S}_1^n{,}M_1{,}Z_{n{-}j{+}2}^n{,}Y_1^{j})\label{eq:remcondy2}\\
{=}&\Theta_{j}^{(1)}{+}\Theta_{n{-}j{+}1}^{(1)}{+}\Phi_{j{+}1}^{(1)}\nonumber\\
&{-}I(Z_{j{+}1}^{n{-}j{+}1};Y_{j}|\mathcal{S}_1^n{,}M_1{,}Z_{n{-}j{+}2}^n{,}Y_1^{j{-}1}){-}
I(Z_{n{-}j{+}1};Y_{j{+}1}^{n{-}j}|\mathcal{S}_1^n{,}M_1{,}Z_{n{-}j{+}2}^n{,}Y_1^{j}).\label{eq:phij1}
\end{align}
The derivations follow the chain rule of mutual information. The inequality in \eqref{eq:remcondy} and \eqref{eq:remcondy2} are due to the fact that removing the condition does not reduce the mutual information (e.g. $I(A;B|C){=}I(A{,}C;B|C){\leq}I(A{,}C;B)$).

When $j{<}\lfloor\frac{n{+}1}{2}\rfloor$, $\Phi_{j}^{(2)}$ is derived as
\begin{align}
\Phi_{j}^{(2)}{=}&
I(M_2;Z_j|M_1{,}\mathcal{S}_1^n{,}Y_1^{j{-}1}{,}Z_{j{+}1}^n){+}I(M_2;Z_{j{+}1}^{n{-}j{+}1}|\mathcal{S}_1^n{,}M_1{,}Y_1^{j{-}1}{,}Z_{n{-}j{+}2}^n)\label{eq:phi21}\\
{=}&I(M_2{,}\hat{\mathcal{S}}_1^n{,}Y_1^{j{-}1}{,}Z_{j{+}1}^n;Z_j|M_1{,}\mathcal{S}_1^n{,}Y_1^{j{-}1}{,}Z_{j{+}1}^n){+}
I(M_2;Z_{n{-}j{+}1}|\mathcal{S}_1^n{,}M_1{,}Y_1^{j{-}1}{,}Z_{n{-}j{+}2}^n)\nonumber\\
&{+}I(M_2;Z_{j{+}1}^{n{-}j}|\mathcal{S}_1^n{,}M_1{,}Y_1^{j{-}1}{,}Z_{n{-}j{+}1}^n)\label{eq:phi22}\\
{\leq}&I(M_2{,}\hat{\mathcal{S}}_1^n{,}Y_1^{j{-}1}{,}Z_{j{+}1}^n;Z_j|M_1{,}\mathcal{S}_1^n{,}Y_1^{j{-}1}{,}Z_{j{+}1}^n){+}
I(M_2{,}Y_{j}^{n{-}j};Z_{n{-}j{+}1}|\mathcal{S}_1^n{,}M_1{,}Y_1^{j{-}1}{,}Z_{n{-}j{+}2}^n)\nonumber\\
&{+}I(M_2{,}Y_j;Z_{j{+}1}^{n{-}j}|\mathcal{S}_1^n{,}M_1{,}Y_1^{j{-}1}{,}Z_{n{-}j{+}1}^n)\label{eq:addytoz}\\
{=}&I(M_2{,}\hat{\mathcal{S}}_1^n{,}Y_1^{j{-}1}{,}Z_{j{+}1}^n;Z_j|M_1{,}\mathcal{S}_1^n{,}Y_1^{j{-}1}{,}Z_{j{+}1}^n)\nonumber\\
&{+}I(M_2;Z_{n{-}j{+}1}|\mathcal{S}_1^n{,}M_1{,}Y_1^{n{-}j}{,}Z_{n{-}j{+}2}^n){+}
I(Y_{j}^{n{-}j};Z_{n{-}j{+}1}|\mathcal{S}_1^n{,}M_1{,}Y_1^{j{-}1}{,}Z_{n{-}j{+}2}^n)\nonumber\\
&{+}I(M_2;Z_{j{+}1}^{n{-}j}|\mathcal{S}_1^n{,}M_1{,}Y_1^{j}{,}Z_{n{-}j{+}1}^n){+}I(Y_j;Z_{j{+}1}^{n{-}j}|\mathcal{S}_1^n{,}M_1{,}Y_1^{j{-}1}{,}Z_{n{-}j{+}1}^n)\label{eq:phi23}\\
{=}&\underbrace{I(M_2{,}\hat{\mathcal{S}}_1^n{,}Y_1^{j{-}1}{,}Z_{j{+}1}^n;Z_j|M_1{,}\mathcal{S}_1^n{,}Y_1^{j{-}1}{,}Z_{j{+}1}^n)}_{\Theta_j^{(2)}}{+}
\underbrace{I(M_2{,}Y_1^{n{-}j}{,}Z_{n{-}j{+}2}^n{,}\hat{\mathcal{S}}_1^n;Z_{n{-}j{+}1}|
\mathcal{S}_1^n{,}M_1{,}Y_1^{n{-}j}{,}Z_{n{-}j{+}2}^n)}_{\Theta_{n{-}j{+}1}^{(2)}}\nonumber\\
&{+}\underbrace{I(M_2;Z_{j{+}1}^{n{-}j}|\mathcal{S}_1^n{,}M_1{,}Y_1^{j}{,}Z_{n{-}j{+}1}^n)}_{\Phi_{j{+}1}^{(2)}}\nonumber\\
&{+}I(Y_{j}^{n{-}j};Z_{n{-}j{+}1}|\mathcal{S}_1^n{,}M_1{,}Y_1^{j{-}1}{,}Z_{n{-}j{+}2}^n){+}
I(Y_j;Z_{j{+}1}^{n{-}j}|\mathcal{S}_1^n{,}M_1{,}Y_1^{j{-}1}{,}Z_{n{-}j{+}1}^n)\label{eq:phi24}\\
{=}&\Theta_j^{(2)}{+}\Theta_{n{-}j{+}1}^{(2)}{+}\Phi_{j{+}1}^{(2)}\nonumber\\
&{+}I(Y_{j}^{n{-}j};Z_{n{-}j{+}1}|\mathcal{S}_1^n{,}M_1{,}Y_1^{j{-}1}{,}Z_{n{-}j{+}2}^n){+}
I(Y_j;Z_{j{+}1}^{n{-}j}|\mathcal{S}_1^n{,}M_1{,}Y_1^{j{-}1}{,}Z_{n{-}j{+}1}^n).\label{eq:phij2}
\end{align}
The derivations follow the chain rule of mutual information. \eqref{eq:addytoz} results from adding $Y_j^{n{-}j}$ and $Y_{j}$ respectively in the last two terms increases the mutual information (e.g.$I(A;B|C){\leq}I(A{,}D;B|C)$, the equality holds if and only if $D$ is deterministic function of $C$.).

Combining \eqref{eq:phij1} and \eqref{eq:phij2} yields
\begin{align}
\Phi_j{=}&\Phi_{j}^{(1)}{+}\Phi_{j}^{(2)}\\
{\leq}&\Theta_{j}^{(1)}{+}\Theta_{n{-}j{+}1}^{(1)}{+}\Phi_{j{+}1}^{(1)}{+}\Theta_j^{(2)}{+}\Theta_{n{-}j{+}1}^{(2)}{+}\Phi_{j{+}1}^{(2)}\nonumber\\
&{+}I(Y_{j}^{n{-}j};Z_{n{-}j{+}1}|\mathcal{S}_1^n{,}M_1{,}Y_1^{j{-}1}{,}Z_{n{-}j{+}2}^n){+}
I(Y_j;Z_{j{+}1}^{n{-}j}|\mathcal{S}_1^n{,}M_1{,}Y_1^{j{-}1}{,}Z_{n{-}j{+}1}^n)\nonumber\\
&{-}I(Z_{j{+}1}^{n{-}j{+}1};Y_{j}|\mathcal{S}_1^n{,}M_1{,}Z_{n{-}j{+}2}^n{,}Y_1^{j{-}1}){-}
I(Z_{n{-}j{+}1};Y_{j{+}1}^{n{-}j}|\mathcal{S}_1^n{,}M_1{,}Z_{n{-}j{+}2}^n{,}Y_1^{j})\\
{=}&\Theta_{j}{+}\Theta_{n{-}j{+}1}{+}\Phi_{j{+}1}\nonumber\\
&{+}I(Y_{j};Z_{n{-}j{+}1}|\mathcal{S}_1^n{,}M_1{,}Y_1^{j{-}1}{,}Z_{n{-}j{+}2}^n){+}
I(Y_{j{+}1}^{n{-}j};Z_{n{-}j{+}1}|\mathcal{S}_1^n{,}M_1{,}Y_1^{j}{,}Z_{n{-}j{+}2}^n)\nonumber\\
&{+}I(Y_j;Z_{j{+}1}^{n{-}j}|\mathcal{S}_1^n{,}M_1{,}Y_1^{j{-}1}{,}Z_{n{-}j{+}1}^n)\nonumber\\
&{-}I(Z_{n{-}j{+}1};Y_{j}|\mathcal{S}_1^n{,}M_1{,}Z_{n{-}j{+}2}^n{,}Y_1^{j{-}1}){-}
I(Z_{j{+}1}^{n{-}j};Y_{j}|\mathcal{S}_1^n{,}M_1{,}Z_{n{-}j{+}1}^n{,}Y_1^{j{-}1})\nonumber\\
&{-}I(Z_{n{-}j{+}1};Y_{j{+}1}^{n{-}j}|\mathcal{S}_1^n{,}M_1{,}Z_{n{-}j{+}2}^n{,}Y_1^{j})\\
{=}&\Theta_{j}{+}\Theta_{n{-}j{+}1}{+}\Phi_{j{+}1}.
\end{align}

\subsection{When $j{=}\lfloor\frac{n{+}1}{2}\rfloor$}
\subsubsection{$n$ is an even number}
In this case, $j{=}\lfloor\frac{n{+}1}{2}\rfloor{=}\frac{n}{2}$. Following \eqref{eq:phi11}, \eqref{eq:remcondy1} and \eqref{eq:remcondy2}, we have
\begin{align}
\Phi_{\frac{n}{2}}^{(1)}{=}&I(M_1{,}Z_{\frac{n}{2}{+}2}^n;Y_{\frac{n}{2}{+}1}|\mathcal{S}_1^n{,}Y_1^{\frac{n}{2}})
{+}I(M_1{,}Z_{\frac{n}{2}{+}2}^n;Y_{\frac{n}{2}}|\mathcal{S}_1^n{,}Y_1^{\frac{n}{2}{-}1})\\
{\leq}&I(M_1{,}Z_{\frac{n}{2}{+}2}^n{,}Y_1^{\frac{n}{2}};Y_{\frac{n}{2}{+}1}|\mathcal{S}_1^n)
{+}I(M_1{,}Z_{\frac{n}{2}{+}2}^n;Y_{\frac{n}{2}}|\mathcal{S}_1^n{,}Y_1^{\frac{n}{2}{-}1})\\
{=}&I(M_1{,}Z_{\frac{n}{2}{+}2}^n{,}Y_1^{\frac{n}{2}};Y_{\frac{n}{2}{+}1}|\mathcal{S}_1^n)
{+}I(M_1{,}Z_{\frac{n}{2}{+}1}^n;Y_{\frac{n}{2}}|\mathcal{S}_1^n{,}Y_1^{\frac{n}{2}{-}1})\nonumber\\
&{-}I(Z_{\frac{n}{2}{+}1};Y_{\frac{n}{2}}|\mathcal{S}_1^n{,}M_1{,}Z_{\frac{n}{2}{+}2}^n{,}Y_1^{\frac{n}{2}{-}1})\\
{\leq}&\underbrace{I(M_1{,}Z_{\frac{n}{2}{+}2}^n{,}Y_1^{\frac{n}{2}}{,}\hat{\mathcal{S}}_1^n;Y_{\frac{n}{2}{+}1}|\mathcal{S}_1^n)}_{\Theta_{\frac{n}{2}{+}1}}
{+}\underbrace{I(M_1{,}Z_{\frac{n}{2}{+}1}^n{,}Y_1^{\frac{n}{2}{-}1}{,}\hat{\mathcal{S}}_1^n;Y_{\frac{n}{2}}|\mathcal{S}_1^n)}_{\Theta_{\frac{n}{2}}}\nonumber\\
&{-}I(Z_{\frac{n}{2}{+}1};Y_{\frac{n}{2}}|\mathcal{S}_1^n{,}M_1{,}Z_{\frac{n}{2}{+}2}^n{,}Y_1^{\frac{n}{2}{-}1})\label{eq:phi2_1}.
\end{align}

$\Phi_{\frac{n}{2}}^{(2)}$ is derived similar as \eqref{eq:phi21}-\eqref{eq:phi24}, namely
\begin{align}
\Phi_{\frac{n}{2}}^{(2)}{=}&I(M_2;Z_\frac{n}{2}|M_1{,}\mathcal{S}_1^n{,}Y_1^{\frac{n}{2}{-}1}{,}Z_{\frac{n}{2}{+}1}^n)
{+}I(M_2;Z_{\frac{n}{2}{+}1}|\mathcal{S}_1^n{,}M_1{,}Y_1^{\frac{n}{2}{-}1}{,}Z_{\frac{n}{2}{+}2}^n)\\
{=}&\underbrace{I(M_2{,}\hat{\mathcal{S}}_1^n{,}Y_1^{\frac{n}{2}{-}1}{,}Z_{\frac{n}{2}{+}1}^n;Z_\frac{n}{2}|M_1{,}\mathcal{S}_1^n{,}Y_1^{\frac{n}{2}{-}1}{,}Z_{\frac{n}{2}{+}1}^n)}_{\Theta_{\frac{n}{2}}}
{+}I(M_2;Z_{\frac{n}{2}{+}1}|\mathcal{S}_1^n{,}M_1{,}Y_1^{\frac{n}{2}{-}1}{,}Z_{\frac{n}{2}{+}2}^n)\\
{\leq}&\Theta_{\frac{n}{2}}{+}
I(M_2{,}Y_{\frac{n}{2}};Z_{\frac{n}{2}{+}1}|\mathcal{S}_1^n{,}M_1{,}Y_1^{\frac{n}{2}{-}1}{,}Z_{\frac{n}{2}{+}2}^n)\\
{=}&\Theta_{\frac{n}{2}}{+}I(M_2;Z_{\frac{n}{2}{+}1}|\mathcal{S}_1^n{,}M_1{,}Y_1^{\frac{n}{2}}{,}Z_{\frac{n}{2}{+}2}^n){+}
I(Y_{\frac{n}{2}};Z_{\frac{n}{2}{+}1}|\mathcal{S}_1^n{,}M_1{,}Y_1^{\frac{n}{2}{-}1}{,}Z_{\frac{n}{2}{+}2}^n)\\
{=}&\Theta_{\frac{n}{2}}{+}\underbrace{I(M_2{,}\hat{\mathcal{S}}_1^n{,}Y_1^{\frac{n}{2}}{,}Z_{\frac{n}{2}{+}2}^n;Z_{\frac{n}{2}{+}1}|\mathcal{S}_1^n{,}M_1{,}Y_1^{\frac{n}{2}}{,}Z_{\frac{n}{2}{+}2}^n)}_{\Theta_{\frac{n}{2}{+}1}}{+}
I(Y_{\frac{n}{2}};Z_{\frac{n}{2}{+}1}|\mathcal{S}_1^n{,}M_1{,}Y_1^{\frac{n}{2}{-}1}{,}Z_{\frac{n}{2}{+}2}^n)\label{eq:phi2_2}
\end{align}

Adding \eqref{eq:phi2_1} and \eqref{eq:phi2_2}, \eqref{eq:recursive2} holds.

\subsubsection{$n$ is an odd number}
In this case, $\lfloor\frac{n{+}1}{2}\rfloor{=}\frac{n{+}1}{2}$. We can rewrite $\Phi_{\frac{n{+}1}{2}}^{(1)}$ and $\Phi_{\frac{n{+}1}{2}}^{(2)}$ as
\begin{align}
\Phi_{\frac{n{+}1}{2}}^{(1)}{=}&I(M_1{,}Z_{\frac{n{+}1}{2}{+}1}^n;Y_\frac{n{+}1}{2}|\mathcal{S}_1^n{,}Y_1^{\frac{n{+}1}{2}{-}1})\\
{=}&\underbrace{I(M_1{,}Z_{\frac{n{+}1}{2}{+}1}^n{,}Y_1^{\frac{n{+}1}{2}{-}1}{,}
\hat{\mathcal{S}}_1^n;Y_\frac{n{+}1}{2}|\mathcal{S}_1^n{,}Y_1^{\frac{n{+}1}{2}{-}1})}_{\Theta_{\frac{n{+}1}{2}}^{(1)}},\label{eq:phi1odd}\\
\Phi_{\frac{n{+}1}{2}}^{(2)}{=}&I(M_2;Z_\frac{n{+}1}{2}|M_1{,}\mathcal{S}_1^n{,}Y_1^{\frac{n{+}1}{2}{-}1}{,}Z_{\frac{n{+}1}{2}{+}1}^n)\\
{=}&\underbrace{I(M_2{,}Y_1^{\frac{n{+}1}{2}{-}1}{,}Z_{\frac{n{+}1}{2}{+}1}^n{,}\hat{\mathcal{S}}_1^n;
Z_\frac{n{+}1}{2}|M_1{,}\mathcal{S}_1^n{,}Y_1^{\frac{n{+}1}{2}{-}1}{,}Z_{\frac{n{+}1}{2}{+}1}^n)}_{\Theta_{\frac{n{+}1}{2}}^{(2)}}.\label{eq:phi2odd}
\end{align}
Consequently, \eqref{eq:recursive2} holds after adding \eqref{eq:phi1odd} and \eqref{eq:phi2odd}. 

\ifCLASSOPTIONcaptionsoff
  \newpage
\fi

\bibliographystyle{IEEEtran}

\bibliography{freq_jrnl}

\end{document}